\documentclass{article}
\usepackage{geometry}
\usepackage{units}
\usepackage{graphicx}
\usepackage{amsmath}
\usepackage{amssymb}
\usepackage{calc}
\usepackage{particles}
\usepackage{subfigure}

\usepackage{changebar}

\newcommand{\hashh}{{\#}}
\newcommand{\updag}{{\ \ensuremath{^\dag}}}
\newcommand{\upddag}{{\ \ensuremath{^\ddag}}}
\newcommand{\updollar}{{\ \ensuremath{^\$}}}

\newcommand{\uphashh}{{\ \ensuremath{^{\text{\hashh}}}}}

\newcommand{\pngb}{\particle{P^a}}

\makeatletter
\newlength{\DS@width}
\newlength{\slash@width}
\newlength{\DS@offset}{}
\DeclareRobustCommand{\DiracSlash}[1]{
  \settowidth{\DS@width}{\ensuremath{#1}}
  \setlength{\DS@offset}{\DS@width/2}\addtolength{\DS@offset}{-\slash@width/2}
  {\makebox[0in][l]{\hspace{\DS@offset}$/$}\ensuremath{#1}}
}
\makeatother

\newcommand{\missingE}{\DiracSlash{\mathrm{E}}}
\newcommand{\branching}[1]{{\ensuremath{\mathrm{BR}({#1})}}}
\newcommand{\crosssec}[1]{{\ensuremath{\mathrm{\sigma}({#1})}}}
\newcommand{\width}[1]{{\ensuremath{\mathrm{\Gamma}({#1})}}}
\newcommand{\NTC}{\ensuremath{N_{\mathrm{TC}}}}
\newcommand{\anom}{\ensuremath{\mathcal{A}_{G_1 G_2}}} 
\newcommand{\anomppp}{\ensuremath{\mathcal{A}_{\photon\photon}}}
\newcommand{\anompzp}{\ensuremath{\mathcal{A}_{\Znaught\photon}}}
\newcommand{\anompzz}{\ensuremath{\mathcal{A}_{\Znaught\Znaught}}}
\newcommand{\anompzG}{\ensuremath{\mathcal{A}_{\Znaught G}}}
\newcommand{\anomppG}{\ensuremath{\mathcal{A}_{\photon G}}}

\newcommand{\Fpzp}{\ensuremath{\mathcal{F}_{\Znaught \photon}^\photon}}
\newcommand{\Fppp}{\ensuremath{\mathcal{F}_{\photon \photon}^\photon}}
\newcommand{\ffrat}{\ensuremath{(\unit[123]{GeV}/f_\pngb)}}

\newcommand{\crossppp}{\ensuremath{\sigma_{\photon\photon}^\photon}}
\newcommand{\crosszpp}{\ensuremath{\sigma_{\Znaught\photon}^\photon}}
\newcommand{\crosszpz}{\ensuremath{\sigma_{\Znaught\photon}^\Znaught}}
\newcommand{\crosszzz}{\ensuremath{\sigma_{\Znaught\Znaught}^\Znaught}}

\newcommand{\swsq}{\ensuremath{s_W^2}}
\newcommand{\cwsq}{\ensuremath{c_W^2}}
\newcommand{\swfour}{\ensuremath{s_W^4}}
\newcommand{\cwfour}{\ensuremath{c_W^4}}
\newcommand{\swsix}{\ensuremath{s_W^6}}
\newcommand{\cwsix}{\ensuremath{c_W^6}}

\newcommand{\sinsqtw}{\ensuremath{\sin^2\theta_W}}
\newcommand{\cossqtw}{\ensuremath{\cos^2\theta_W}}
\newcommand{\sinfourtw}{\ensuremath{\sin^4\theta_W}}
\newcommand{\cosfourtw}{\ensuremath{\cos^4\theta_W}}
\newcommand{\jetj}{\ensuremath{\mathrm{j}}}
\newcommand{\TESLA}{TESLA}
\newcommand{\GigaZ}{GigaZ}

\newcommand{\psibar}{\ensuremath{\overline{\psi}}}
\newcommand{\gammafive}{\ensuremath{\gamma_5}}

\newcommand{\technipi}{\particle{\pi_T^0}}
\newcommand{\technipiprime}{\ensuremath{\particle{\pi_T^{0\prime}}}}
\newcommand{\Ltechni}{\particle{L}}
\newcommand{\Qtechni}{\particle{Q}}
\newcommand{\Dtechni}{\particle{D}}
\newcommand{\Ltechnibar}{\ensuremath{\overline{\Ltechni}}}
\newcommand{\Qtechnibar}{\ensuremath{\overline{\Qtechni}}}
\newcommand{\Dtechnibar}{\ensuremath{\overline{\Dtechni}}}
\newcommand{\Ntechni}{\particle{N}}
\newcommand{\Etechni}{\particle{E}}
\newcommand{\Ntechnibar}{\ensuremath{\overline{\Ntechni}}}
\newcommand{\Etechnibar}{\ensuremath{\overline{\Etechni}}}
\newcommand{\TUtechni}{\particle{T_U}}
\newcommand{\TDtechni}{\particle{T_D}}
\newcommand{\TUtechnibar}{\ensuremath{\overline{\particle{T}}_\particle{U}}}
\newcommand{\TDtechnibar}{\ensuremath{\overline{\particle{T}}_\particle{D}}}

\newcommand{\neutralinoone}{\ensuremath{\tilde{\chi}^0_1}}
\newcommand{\neutralinotwo}{\ensuremath{\tilde{\chi}^0_2}}

\newcommand{\Pone}{\particle{P^1}}
\newcommand{\Pthree}{\particle{P^3}}
\newcommand{\Pfivem}{\particle{P^5_-}}
\newcommand{\Pfivep}{\particle{P^5_+}}

\newcommand{\PEtechni}{\particle{P_\Etechni}}
\newcommand{\PNtechni}{\particle{P_\Ntechni}}

\newcommand{\Nltechni}{\ensuremath{\particle{N}_l}}
\newcommand{\Nltechnibar}{\ensuremath{\overline{\particle{N}}_l}}
\newcommand{\Eltechni}{\ensuremath{\particle{E}_l}}
\newcommand{\Eltechnibar}{\ensuremath{\overline{\particle{E}}_l}}
\newcommand{\PthreeL}{\particle{P^3_\Ltechni}}

\newcommand{\ffbar}{\ensuremath{{\particle{f}\antiparticle{f}}}}

\newcommand{\SUNTC}{\group{SU}{\NTC}}
\newcommand{\gTC}{\ensuremath{g_{\mathrm{TC}}}}
\newcommand{\betaTC}{\ensuremath{\beta_{\mathrm{TC}}}}

\DeclareMathOperator{\trace}{Tr}
\DeclareMathOperator{\diag}{diag}

\makeatletter
\@addtoreset{equation}{section}
\makeatother

\makeatletter
\renewcommand{\@thesubfigure}{(\alph{subfigure})\space}
\renewcommand{\p@subfigure}{}
\makeatother

\title{Composite Scalars at \LEP: Constraining
  Technicolor Theories}
\author{Kevin R. Lynch\thanks{krlynch@bu.edu}$^{\text{ , 1}}$,
  Elizabeth H. Simmons\thanks{simmons@bu.edu}$^{\text{ , 1, 2}}$\\
  \\
  $^1$ Department of Physics, Boston University, \\
  590 Commonwealth Avenue, Boston MA  02215\\
  \\
  $^2$ Radcliffe Institute for Advanced Study, Harvard University,\\
  34 Concord Avenue, Cambridge, MA  02138\\
  and\\
  Jefferson Physical Laboratory, Harvard University\\
  Cambridge, MA  02138}
\date{March 14, 2001}

\begin{document}

\begin{titlepage}
\maketitle
\thispagestyle{empty}

  \begin{picture}(0,0)(0,0)
    \put(400,300){BUHEP-00-19}
    \put(400,290){HUPT-01/A005}
    \put(400,280){hep-ph/0012256}
  \end{picture}
  \vspace{24pt}

\begin{abstract} \LEPI\ and \LEPII\ data can be used to constrain
  technicolor models with light, neutral pseudo-Nambu-Goldstone bosons,
  \pngb.  We use published limits on branching ratios and cross sections for
  final states with photons, large missing energy, jet pairs, or \bbbar\
  pairs to constrain the anomalous $\pngb\Znaught\Znaught$,
  $\pngb\Znaught\photon$, and $\pngb\photon\photon$ couplings.  From these
  results, we derive bounds on the size of the technicolor gauge group and
  the number of technifermion doublets in models such as Low-scale
  Technicolor.
\end{abstract}

\end{titlepage}

\section{Introduction}
\label{sec:introduction}

Although the scale of electroweak symmetry breaking is well
established, the mechanism of that breaking is still unknown.  Data
collected at \LEP\ over the last twelve years, however, have provided
many constraints on the properties of that mechanism.  In this paper,
we consider what the \LEP\ data reveal about non-minimal technicolor
models.  In particular, we explore how the limits on rare processes
constrain technicolor models with neutral pseudo-Nambu-Goldstone
bosons (PNGBs), \pngb, which couple, through an anomaly, to the
neutral electroweak bosons.  PNGBs lighter than the \Znaught\ can be
produced at the \Znaught\ pole through the decays $\Znaught\to
\photon\pngb$ or $\Znaught \to \Zstar\pngb$, while heavier PNGBs can
be produced through a number of processes at the higher energies found
at \LEPII.  Depending on the details of the specific model, the final
state following PNGB decay may include jets, photons, or missing
energy, providing striking signatures.

Our analysis is not the first to consider these processes~\cite{am-lr,
  lr-ehs, gr-ehs, vl-ps, rc-etal}.  Since the work of
Reference~\cite{gr-ehs}, however, the \LEP\ collaborations have published
new analyses using additional \LEPI\ data \cite{L3photons,
  DELPHImissing, L3hadron, OPALhadron, OPALzstarp}, allowing stronger
limits to be placed on the $\pngb\Znaught\photon$ couplings.
Furthermore, improvements in the resolution of photon energy
measurements allow the limits to be extended to larger PNGB masses.
The quality of the final \LEPI\ data are such that, contrary to
previous expectations, bounds can even be placed on the
$\pngb\Znaught\Znaught$ coupling.  Finally, some of the data collected
at \LEPII\ has been analyzed \cite{OPAL-LEPII, DELPHI-LEPII, L3-LEPII,
  DELPHI-missing-LEPII, ALEPH-hz-LEPII, L3-hz-LEPII} and provides a
means both to search for heavier PNGBs and to place bounds on the
$\pngb\photon\photon$ couplings for the first time.  The constraints
on modern, non-minimal technicolor models derived from all of these
coupling bounds are phenomenologically interesting.
  
In the next section, we review the production and primary decay
mechanisms for technicolor PNGBs at \LEP\ through the anomalies.  In
Section~\ref{sec:lepi}, we analyze the limits on the anomalous
$\pngb\Znaught\photon$ and $\pngb\Znaught\Znaught$ couplings that can
be derived from published analyses of \LEPI\ data.  In
Section~\ref{sec:lepii} we likewise derive bounds on the anomalous
$\pngb\photon\photon$, $\pngb\Znaught\photon$ and
$\pngb\Znaught\Znaught$ couplings from published analyses of \LEPII\ 
data.  Section~\ref{sec:combined} compiles the strongest results for
each anomalous coupling and \pngb\ decay mode.  In
Section~\ref{sec:implications} we determine what these results imply
for various technicolor scenarios.  We present our conclusions and
thoughts about the future in Section~\ref{sec:conclusions}.

\section{Production and decay of \pngb}
\label{sec:equations}

\begin{figure}
\begin{center}
  \subfigure[$\epem\to\pngb\photon$]{\includegraphics[width=
    (\textwidth-1in)/3]{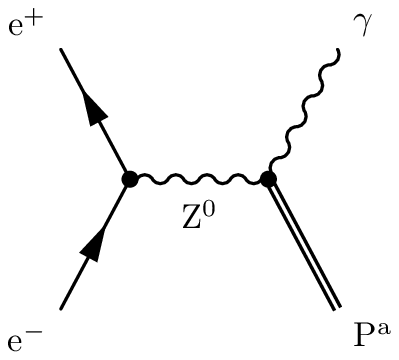} \label{fig:fds:lepi:PaP}} \qquad\qquad
  \subfigure[$\epem\to\pngb\Zstar$]{\includegraphics[width=
    (\textwidth-1in)/3]{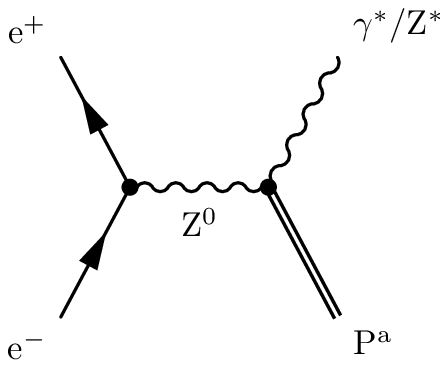} \label{fig:fds:lepi:PaZs}}
\caption{Primary production mechanisms of PNGBs at \LEPI.  At \LEPI,
  processes with an intermediate, on-shell \Znaught\ dominate the
  cross section for any PNGB production processes.  The left hand
  diagram is the relevant one for processes with mono-energetic,
  hard photons, plus the \pngb\ decay products in the final state;
  these states give clean access to the $\pngb\Znaught\photon$
  coupling.  The right hand diagram is the relevant one for 
  processes with four particles in the final state, and will generally give
   access to both the $\pngb\Znaught\photon$ and
  $\pngb\Znaught\Znaught$ couplings.}
\label{fig:fds:lepi}
\end{center}
\end{figure}

\begin{figure}
\begin{center}
  \subfigure[$\epem \to \pngb\photon$]{\includegraphics[width=
    (\textwidth-1in)/3]{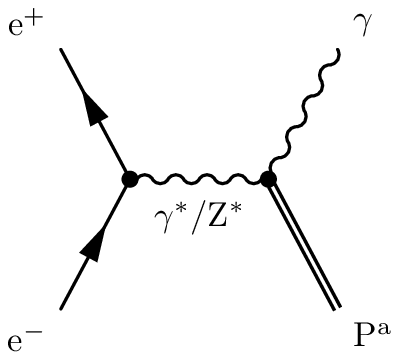} \label{fig:fds:lepii:PaP}} 
  \qquad\qquad
  \subfigure[$\epem \to \pngb\Znaught$]{\includegraphics[width=
    (\textwidth-1in)/3]{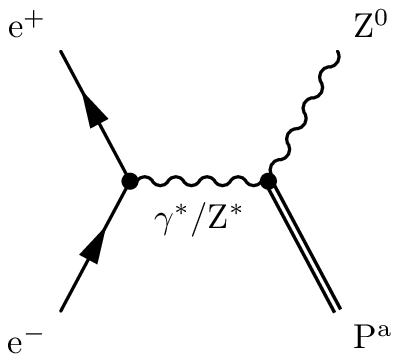} \label{fig:fds:lepii:PaZ}}\\
  \subfigure[$\epem \to
  \pngb(\photon^*/\Zstar)$]{\includegraphics[width=
    (\textwidth-1in)/3]{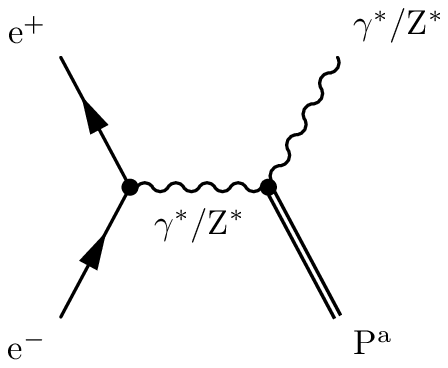} \label{fig:fds:lepii:PaPsZs}}
  \qquad\qquad
  \subfigure[$\epem \to \pngb\epem$]{\includegraphics[width=
    (\textwidth-1in)/3]{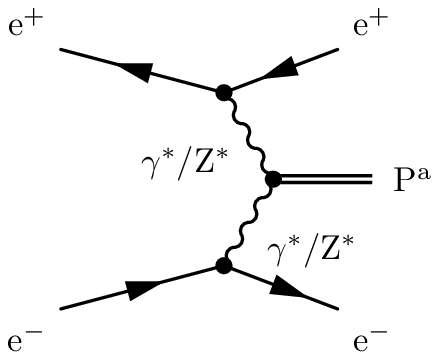} \label{fig:fds:lepii:Paepem}}
\caption{Primary production mechanisms of PNGBs at \LEPII.  
  The first type of process is $s$-channel production via an intermediate,
  off-shell photon or \Znaught.  The diagram at upper left is the relevant
  one for processes with a hard, mono-energetic photon plus the decay
  products of the \pngb\ in the final state, and gives access to both the
  $\pngb\photon\photon$ and $\pngb\Znaught\photon$ couplings.  The diagram at
  upper right is the relevant one for processes with a real \Znaught\ plus
  the decay products of the \pngb\ in the final state, and gives access to
  both the $\pngb\Znaught\photon$ and $\pngb\Znaught\Znaught$ couplings.  The
  diagram at lower left is also, in principle, of relevance at \LEPII, and
  would give access to all of the various couplings of electroweak gauge
  bosons to PNGBs; however, these processes are much more difficult to
  analyze, and are not studied here.  Finally, the diagram at lower right
  would, in principle, give access to all of the anomalous couplings of the
  \pngb; however, kinematics strongly favors the process with intermediate
  photons, so that only the $\pngb\photon\photon$ coupling is accessible.}
\label{fig:fds:lepii}
\end{center}
\end{figure}

At LEP I, a light neutral PNGB, \pngb, with $M_\pngb < M_\Znaught$ is
primarily produced \cite{am-lr, lr-ehs, gr-ehs} through an anomalous
coupling to the \Znaught\ boson and either a photon ($\Znaught \to
\photon\pngb$) or a second, off-shell \Znaught\ boson ($\Znaught \to
\Zstar\pngb$).  At the higher center of mass energies of \LEPII,
PNGBs over a wider range of masses can be produced through $s$-channel
$\photon^*/\Zstar$ exchange and through a $2 \to 3$ production
mechanism~\cite{vl-ps, rc-etal}.  For reference, we provide Feynman
diagrams in Figures~\ref{fig:fds:lepi} and~\ref{fig:fds:lepii}.

The anomalous coupling between the
PNGB and the gauge bosons $\particle{G}_1$ and $\particle{G}_2$ is
given, in a model with technicolor group \SUNTC, by an expression analogous
to that for the QCD pion~\cite{sd-sr-glk, je-etal, bh}
\begin{equation}
  \NTC\anom \frac{g_1 g_2}{2 \pi^2 f_\pngb}
  \epsilon_{\mu\nu\lambda\sigma} k^\mu_1 k^\nu_2 \varepsilon^\lambda_1
  \varepsilon^\sigma_2\ ,
\end{equation}
where \NTC\ is the number of technicolors, \anom\ is the anomaly
factor (discussed further below), the $g_i$ are the gauge couplings of
the gauge bosons, and the $k_i$ and $\varepsilon_i$ are the
four-momenta and polarizations of the gauge bosons.  The \pngb\ decay
constant, $f_\pngb$, which corresponds to the QCD pion decay constant,
$f_\pi$, is given by~\cite{lr-ehs}
\begin{equation}
  f^2_\pngb = \frac{v^2}{2 \trace\left[\left(T_L - T_R\right)^2\right]}\ 
  , 
  \label{eqn:ftechnifirst}
\end{equation}
where $v =$ \unit[246]{GeV} is the weak scale, and $T_L$ ($T_R$) is
the charged weak generator associated with the left-handed
(right-handed) technifermions that comprise the PNGB.  In the case of
left-handed electroweak doublet techniquarks, \Qtechni\ (which are
$\SUthree_C$ triplets), and technileptons, \Ltechni\ (which are
$\SUthree_C$ singlets), the above expression reduces to
\begin{equation}
  \label{eqn:ftechnisecond}
  f_\pngb = \frac{v}{\sqrt{3 N_\Qtechni + N_\Ltechni}} ,
\end{equation}
where the $N_i$ are the number of such electroweak doublets in the
model.  Equation~\ref{eqn:ftechnifirst} is only valid in the
limit of small isospin breaking in the technifermion sector (in
Section~\ref{sec:onefamily} we consider the consequences of a particular
case of large isospin breaking).

The rate of PNGB production at the $\Znaught$ pole has previously been
reported in the literature; the cross section for production at
$\sqrt{s} = M_\Znaught$ is~\cite{pdb}
\begin{equation}
\crosssec{\epem \to \Znaught \to \pngb\particle{X}} = \crosssec{\epem \to
\Znaught}\, \branching{\Znaught \to \pngb\particle{X}} =
\frac{12\pi}{M_\Znaught^2}\, \branching{\Znaught \to \epem}\,
\branching{\Znaught \to \pngb\particle{X}}\ .
\end{equation}  Production in combination with a
photon~\cite{am-lr} has a width of
\begin{equation}
  \label{eqn:pzpwidth}
  \width{\Znaught \to \photon\pngb} = \unit[2.3\times10^{-5}]{GeV}
  \left(\frac{\unit[123]{GeV}}{f_\pngb}\right)^2 
  \left(1 - \frac{M_\pngb^2}{M_\Znaught^2}\right)^3
  \left(\NTC\anompzp\right)^2\ . 
\end{equation}
Since the measured \Znaught\ width is $\Zwidth = \unit[2.490]{GeV}$
\cite{pdb}, we expect this branching ratio to be of order $10^{-5}$.  The
resulting final states contain a hard mono-energetic photon and the
decay products of the \pngb.  Production in combination with an off-shell
\Znaught\ will be harder to observe.  An upper bound on the decay width of
the process $\Znaught \to \Zstar\pngb \to \pngb\ffbar$ is given
in~\cite{lr-ehs} by\footnote{We have corrected here a slight error in the
  numerical coefficient of the formula as it appears in~\cite{lr-ehs}.  We
  have also included the color factor, $C_f$, which was omitted there.}
\begin{multline}
  \label{eqn:pzzwidth}
  \width{\Znaught \to \pngb \ffbar} <
  \unit[7.6\times10^{-7}]{GeV}
  \left(\frac{\unit[123]{GeV}}{f_\pngb}\right)^2 C_f \left(g_L^2 +
    g_R^2\right)\left(\NTC\anompzz\right)^2 \left(\frac{M_\Znaught^2 -
      M_\pngb^2}{M_\Znaught^6}\right) \times \\
  \left[\left(M_\Znaught - M_\pngb\right)^2 \left(M_\Znaught^2 - 6
      M_\Znaught M_\pngb - 5 M_\pngb^2\right) - 2 M_\pngb^2 \left(6
      M_\Znaught^2 - M_\pngb^2\right) \log \left(\frac{2 M_\Znaught
        M_\pngb - M_\pngb^2}{M_\Znaught^2}\right)\right]\ ,
\end{multline}
where $C_f$ is a color factor of 1 for leptons and 3 for quarks, and
$g_L$ ($g_R$) is the left-handed (right-handed) coupling of the
fermion \particle{f}\ to the \Znaught.  We expect branching ratios of
order $10^{-7}$ to $10^{-6}$, depending on the process of interest.

The production cross section for a PNGB along with an electroweak
gauge boson, $G$, at the higher center of mass energies of \LEPII\ can
be calculated, and has also been reported in the
literature~\cite{lr-ehs}. If $\sqrt{s} - M_\pngb > M_G$, it is possible
to produce a PNGB in association with either an on-shell \Znaught\ 
boson or a photon; the cross section well off the \Znaught\ peak is
given by
\begin{multline}
  \crosssec{\epem \to \pngb G} = \frac{\alphaem^3 \NTC^2}{6\pi^2
    f_\pngb^2 s}\, F(G)\, \lambda(s,M_G^2,M_\pngb^2)^{3/2}\\
  \times \left[ \frac{\anomppG^2}{s^2} +
    \frac{\anomppG\anompzG(1-4\swsq)}{2 \swsq \cwsq s (s -
      M_\Znaught^2)} + \frac{\anompzG^2(1-4\swsq+8\swfour)}{8 \swfour
      \cwfour (s - M_\Znaught^2)^2} \right]\ ,
\label{eqn:epemPaG}
\end{multline}
where $G$ is either the on-shell \Znaught\ or \photon\ in the final
state, $\swfour = \sinfourtw$, $\cwfour = \cosfourtw$, $\lambda(a,b,c)
= a^2 + b^2 + c^2 - 2 ab -2 ac -2 bc$ and
\begin{equation}
F(G) = \begin{cases} 1 & G = \photon\\ \frac{1}{\swsq \cwsq} & G =
  \Znaught\ .
\end{cases}
\end{equation} 
In both cases, the first term is the photon exchange contribution, the
third term is the \Znaught\ exchange contribution, while the second
term is the $\Znaught\photon$ interference term (see
Figure~\ref{fig:fds:lepii}).  Since $\swsq \approx 0.23$, the
interference contribution is generally negligible compared to the
direct contributions.

The model-dependent value of the anomaly factor for the
$\pngb G_1 G_2$ coupling which appears in those branching
ratios is given by~\cite{sd-sr-glk, je-etal, bh}
\begin{equation}
  4 \anom = \trace\left[T^a \left(
      T_1 T_2 + T_2 T_1\right)_L\right] + \trace\left[T^a \left( T_1 T_2
      + T_2 T_1\right)_R\right]\ ,
\end{equation}
where $T^a$ is the generator of the axial current associated with
\pngb, the $T_i$ are the generators associated with the gauge boson
$G_i$, and the subscripts $L$ and $R$ denote the left and right handed
technifermion components that comprise \pngb.  The axial currents are
defined as usual,
\begin{equation}
  j^{\mu a}_5 = \psibar \gamma^\mu \gammafive T^a \psi
\end{equation}
and the generators, $T^a$, are normalized such that
\begin{equation}
  \label{eqn:normalization}
  \trace{\left(T^a T^b\right)} = \frac{1}{2}\delta^{ab}\ .
\end{equation}
For the three cases with neutral electroweak gauge bosons, the anomaly
factors are~\cite{am-lr} 
\begin{gather}
  \anomppp = \trace\left[T^a Q^2\right] \label{eqn:anomppp}\\
  \anompzp = \frac{1}{2} \trace\left[T^a \left(T_{3L} + T_{3R} -
      2Q\sinsqtw\right)Q\right] \label{eqn:anompzp}\\
  \anompzz = \frac{1}{2} \trace \left[T^a \left(\left(T_{3L} - Q
        \sinsqtw\right)^2 + \left(T_{3R} - Q
        \sinsqtw\right)^2\right)\right] \label{eqn:anompzz}\ .
\end{gather}
We will explicitly evaluate the anomaly factors for a variety of models in
Section~\ref{sec:implications}. 

Our analyses will consider all of the dominant decay modes for the
produced PNGBs.  These fall into three classes:
\begin{enumerate}
\item In models where $\anomppp \neq 0$, the PNGB may decay through the
  anomaly to a pair of photons at a rate~\cite{lr-ehs} 
  \begin{equation}
    \width{\pngb \to \photon\photon} =
    \left(\frac{\NTC\anomppp}{f_\pngb} \right)^2 \frac{\alphaem^2}{8
      \pi^3} M_\pngb^3\ .
    \label{eqn:PaPPwidth}
  \end{equation}
  Even for large $M_\pngb$, this decay width is very narrow; for
  example, with $M_\pngb = M_\Znaught$ and $f_\pngb =
  \unit[123]{GeV}$, we find $\width{\pngb \to \photon\photon} \approx
  (\NTC\anomppp)^2 \times \unit[10^{-1}]{keV}$.
\item The PNGB may decay invisibly into neutrinos or other long-lived
  neutral particles.  Alternatively, the PNGB may be long-lived and
  escape the detector. In either case, \pngb\ will manifest as missing
  energy.
\item The PNGB may decay into hadrons.  This may arise through decays
  into \qantiq\ pairs, with \bantib\ being of particular interest in
  some models. Alternatively, PNGBs comprised of colored
  technifermions may decay into gluon pairs.  If no flavor tagging is
  employed in the experimental analysis, limits on hadronic decays of
  the PNGB are assumed to apply equally well to quark and gluon decay
  modes.

\end{enumerate}
Current experimental data provide bounds on all of these processes.

\section{Limits from \LEPI}
\label{sec:lepi}

In this section we explore the limits that can be obtained on the anomaly
factors \anompzp\ and \anompzz\ from published \LEPI\ data~\cite{L3photons,
  DELPHImissing, L3hadron, OPALhadron, OPALzstarp}, collected at $\sqrt{s} =
M_\Znaught$.  We do so for a number of possible decay modes of the \pngb.
The relevant Feynman diagrams are displayed in Figure~\ref{fig:fds:lepi}.

\subsection{Limits on $\NTC\anompzp$}
\label{sec:pzplimits}

For a \Znaught\ produced at rest and undergoing the two-body decay
$\Znaught \to \photon\pngb$ (Figure~\ref{fig:fds:lepi:PaP}),
energy-momentum conservation fixes the photon energy to
be~\cite{lr-ehs}
\begin{equation}
  E_\photon = \frac{M^2_\Znaught - M^2_\pngb}{2 M_\Znaught}\ .
\end{equation}
This provides a striking set of signatures.  We will now use \LEPI\
data on final states that include at least one hard photon to derive
limits on \NTC\anompzp.

\subsubsection{$\Znaught \to \photon\pngb \to \photon\photon\photon$}
\label{sec:pzp:ppp}

\begin{figure}
  \begin{center}
    \includegraphics[width=(\textwidth-1in)/2]{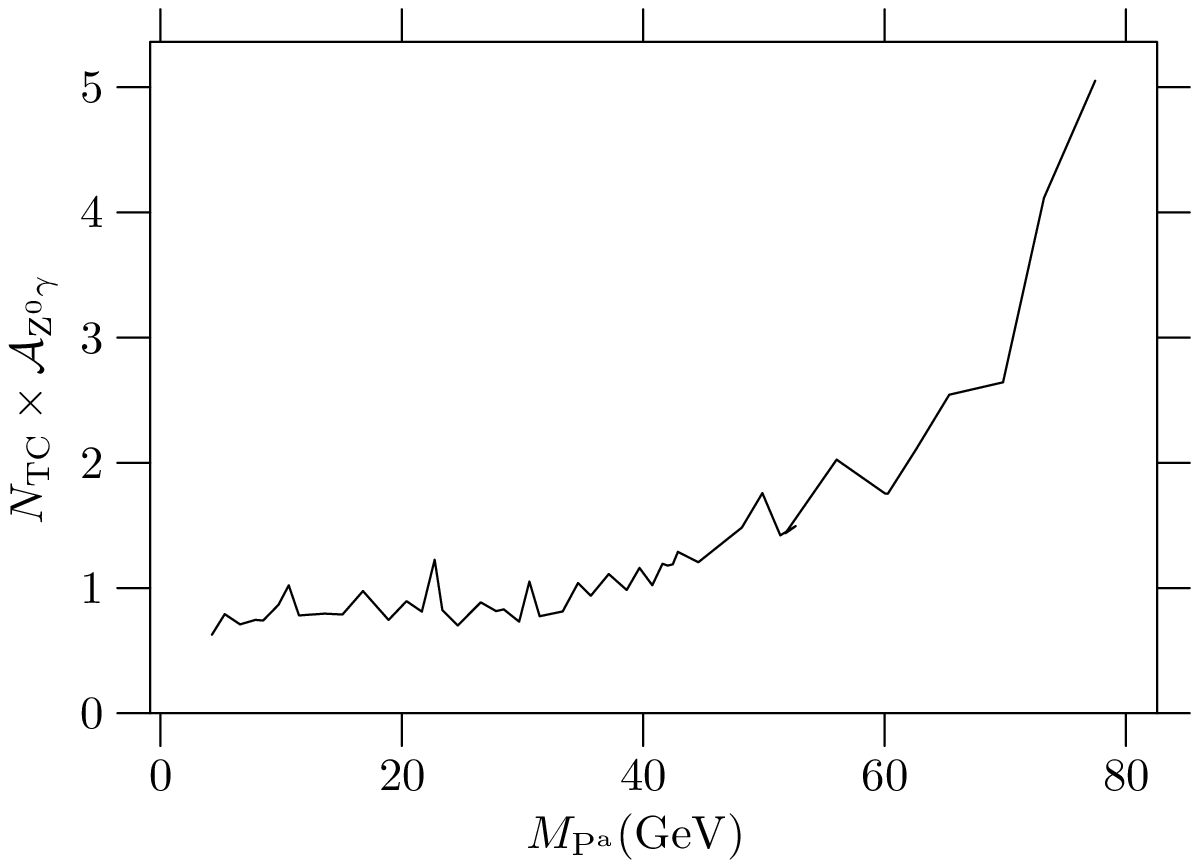}
    \caption{\small Upper limits at 95\% c.l.\ on 
      $\NTC\anompzp\ffrat$ from the process $\Znaught \to \photon\pngb \to
      \photon\photon\photon$.  Our results are derived from an
      \Lthree\ analysis~\cite{L3photons} assuming the PNGB has
      essentially zero width.  Fluctuations in the curves arise from
      fluctuations in the data.}
    \label{fig:LEPI:L3photons}
  \end{center}
\end{figure}

If the PNGB decays dominantly to photons, a final state with three
hard photons results (Figure~\ref{fig:fds:lepi:PaP} with
$\pngb\to\photon\photon$).  The \Lthree\ collaboration has published
limits on the production of a narrow resonance, \particle{X}, decaying
to photons, based on $\unit[65.8]{pb^{-1}}$ of data collected on and
near the \Znaught\ pole~\cite{L3photons}. They find no evidence for a
new resonance, and place 95\% c.l.\ upper limits on the branching
ratio $\branching{\Znaught \to
  \photon\particle{X}}\branching{\particle{X} \to \photon\photon}$ as
a function of $M_\particle{X}$.  For $\unit[3]{GeV} < M_\particle{X} <
\unit[89]{GeV}$, they find $\branching{\Znaught\to
  \photon\particle{X}} \branching{\particle{X} \to \photon\photon} <
1.3 \times 10^{-5}$.

Using Equation~\ref{eqn:pzpwidth}, we translate these data into upper
bounds on $\NTC\anompzp$.  Assuming $\branching{\pngb \to
  \photon\photon} \approx 1$ and $f_\pngb = \unit[123]{GeV}$, we find
$\NTC\anompzp < 0.5-2$ for PNGB masses below \unit[60]{GeV}.  Above
\unit[60]{GeV}, the data become rapidly less constraining (see
Figure~\ref{fig:LEPI:L3photons}).  These limits are a factor of two
stronger than those in Reference~\cite{gr-ehs}.

\subsubsection{$\Znaught \to \photon\pngb \to \photon \missingE$}
\label{sec:pzp:pmissing}

\begin{figure}
\begin{center}
\includegraphics[width=(\textwidth-1in)/2]{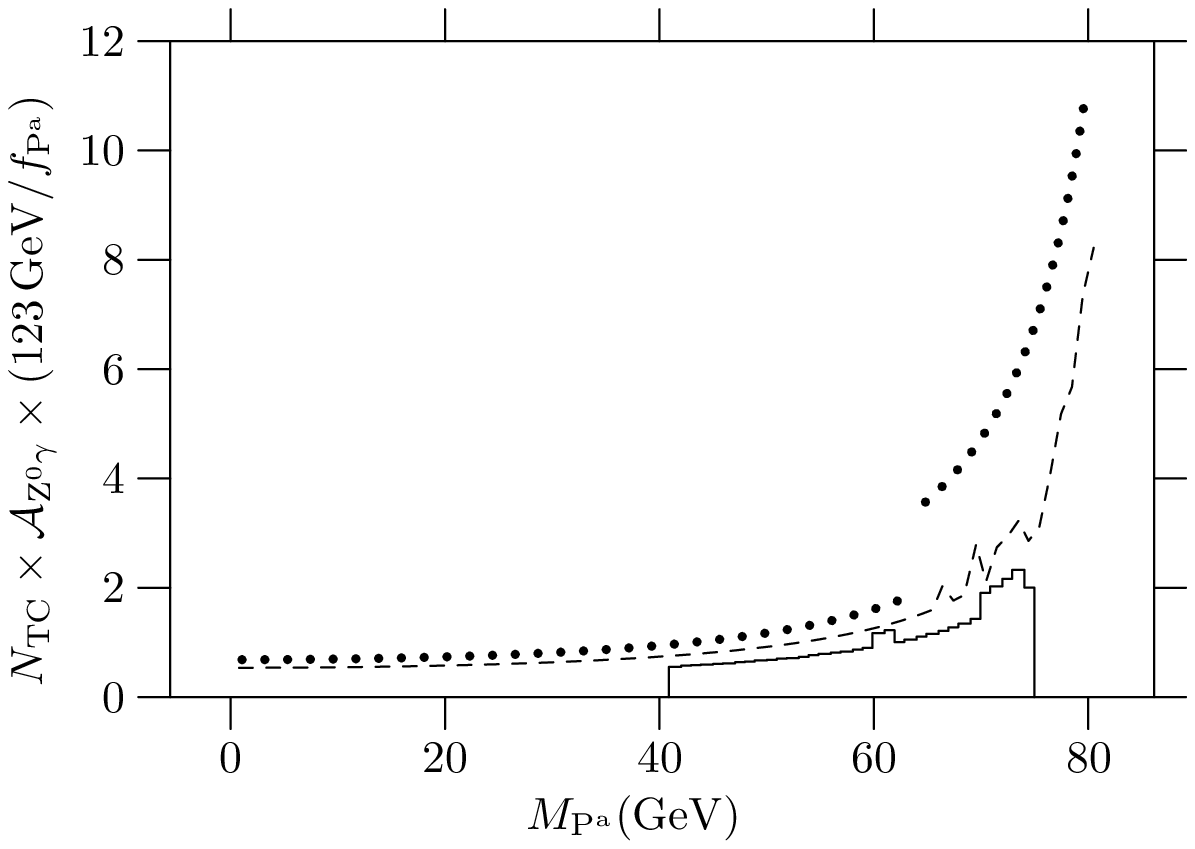}
\caption{\small Upper limits at 95\% c.l.\ on $\NTC\anompzp\ffrat$ from the process $\Znaught
  \to \photon\pngb \to \photon\missingE$.  The dashed line
  corresponds to the results we derived from \DELPHI\ 
  data~\cite{DELPHImissing}.  The dotted curves show the results
  derived from \OPAL\ data~\cite{OPALold} in Reference~\cite{gr-ehs};
  \OPAL\ performed separate searches for scalars with masses below 80
  and 60 GeV.  The solid line shows limits extracted from \Lthree\ 
  data~\cite{L3old} in Reference~\cite{gr-ehs}.}
\label{fig:LEPI:DELPHImissing} \end{center}
\end{figure}

If the predominant decays of the PNGB are invisible, or if it escapes the
detector before decaying, then we expect a final state with one hard photon
and missing energy (Figure~\ref{fig:fds:lepi:PaP} with $\pngb\to \missingE$).
The \DELPHI\ collaboration has searched for anomalous single photon events,
in $\unit[67.6]{pb^{-1}}$ of data collected on and near the \Znaught\ 
pole~\cite{DELPHImissing}. They derive 95\% c.l.\ upper limits on the
production cross section, $\sigma_\particle{X}$, of a narrow
($\Gamma_\particle{X} < \unit[2]{GeV}$) invisible particle \particle{X}\ 
produced in association with a single hard photon, with the photon in the
angular range $\vert\cos\theta\vert < 0.7$ relative to the beamline.  For
$M_\particle{X} < M_\Znaught$, \DELPHI\ provides limits on
$\sigma_\particle{X}$ as a function of $M_\particle{X}$; the upper limit
never exceeds \unit[0.1]{pb}.

Since the \Znaught\ decay is isotropic, we can scale our predictions
to reflect the \DELPHI\ angular coverage.  If we assume that \pngb\ is
always invisible and $f_\pngb = \unit[123]{GeV}$, then using
Equation~\ref{eqn:pzpwidth}, we can derive limits on
$\branching{\Znaught \to \photon\pngb}$, and, hence, $\NTC\anompzp$.
We find $\NTC\anompzp < 0.5 - 1.2$ for \pngb\ masses below
\unit[60]{GeV}; the limits weaken at higher masses.  The limits we
obtain here are stronger than those based on the \OPAL~\cite{OPALold}
data in Reference~\cite{gr-ehs} and cover a larger mass range than those
based on the \Lthree~\cite{L3old} data in Reference~\cite{gr-ehs}. In the
mass range $\unit[40]{GeV} < M_\pngb < \unit[75]{GeV}$ where data from
all three experiments exist, the data from \Lthree\ give the
strongest bounds.  We plot our results in
Figure~\ref{fig:LEPI:DELPHImissing}, along with those of
Reference~\cite{gr-ehs}.

\OPAL\ has also published more recent results on $\photon\missingE$
events, based on $\unit[160]{pb^{-1}}$ of data collected near the
\Znaught\ pole~\cite{OPALzstarp}.  However, since they present this data
as limits on the branching ratios of heavy neutralinos to light
neutralinos and photons via $\Znaught \to \neutralinotwo\neutralinoone
\to \neutralinoone\neutralinoone\photon$, we can not use their results
to constrain $\NTC\anompzp$.

\subsubsection{$\Znaught \to \photon\pngb \to \photon\; \jet\; \jet$}
\label{sec:pzp:pjj}

\begin{figure}
  \begin{center}
    \includegraphics[width=(\textwidth-1in)/2]{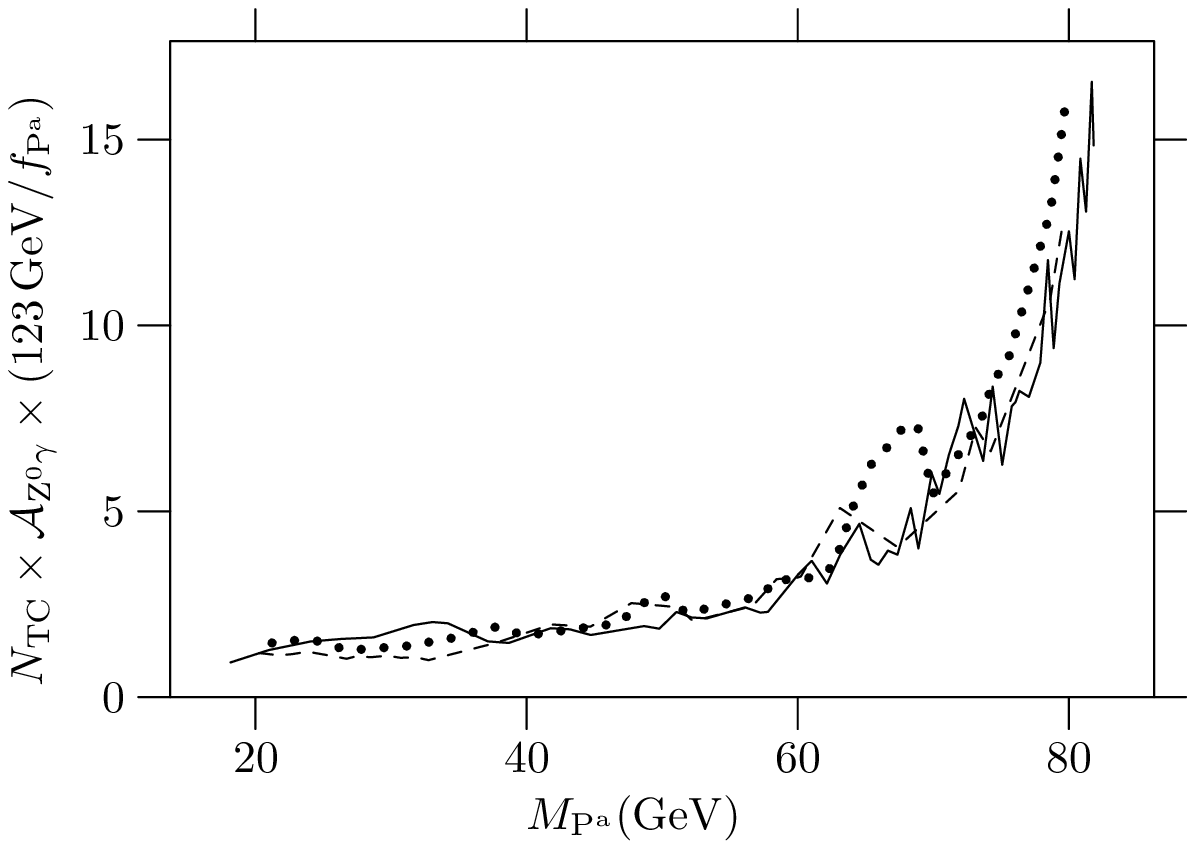}
    \caption{\small Upper limits at 95\% c.l.\ on $\NTC\anompzp\ffrat$ from  $\Znaught \to
      \photon\pngb \to \photon\qantiq$.  We derived the dotted
      (dashed) curve from an \OPAL~\cite{OPALhadron} bound that assumes
      the new scalar decays to \qantiq\ (\bantib).  The solid curve
      comes from \Lthree~\cite{L3hadron} limits for scalar decays to
      hadrons.  Fluctuations in the curves arise from fluctuations in
      the data.}
    \label{fig:LEPI:combined-pjj}
  \end{center}
\end{figure}

If the dominant decay mode of the PNGB is hadronic, a final state with
one hard photon and a pair of jets is expected
(Figure~\ref{fig:fds:lepi:PaP} with $\pngb\to \jet\,\jet$).  Both the
\OPAL\ and \Lthree\ collaborations have published limits on this
process.

\OPAL\ has searched for new, narrow particles decaying to hadrons with an
associated hard photon in $\unit[140]{pb^{-1}}$ of \Znaught\ pole
data~\cite{OPALhadron}. They present two sets of relevant limits: a search for
a scalar resonance, \particle{S^0}, which decays hadronically, and a search
assuming that \particle{S^0} decays predominantly to \bantib.  They find no
evidence for production in either mode, and place 95\% c.l.\ upper limits on
the product of branching ratios, $\branching{\Znaught \to
  \photon\particle{S^0}}\, \branching{\particle{S^0} \to \qantiq}$ as a
function of $M_{\particle{S^0}}$.  For $\unit[20]{GeV} < M_\particle{S^0} <
\unit[80]{GeV}$, the limit always satisfies $\branching{\Znaught \to
  \photon\particle{S^0}}\, \branching{\particle{S^0} \to \qantiq} <
2\times10^{-5}$.  Using Equation~\ref{eqn:pzpwidth}, we translate these
limits into upper bounds on $\NTC\anompzp$, assuming that
$f_\pngb=\unit[123]{GeV}$.  Both sets of data provide limits $\NTC\anompzp <
1-3$ for PNGB masses below \unit[60]{GeV}, and $\NTC\anompzp < 10-15$ for
PNGB masses below \unit[80]{GeV}.

The \Lthree\ collaboration has also searched for new, narrow scalar
particles, \particle{H^0}, decaying to hadrons with an associated hard
photon in $\unit[96.8]{pb^{-1}}$ of data collected at the \Znaught\ 
pole~\cite{L3hadron}. They find no evidence for a new particle, and
place 95\% c.l.\ upper limits on the the cross section for the
process $\Znaught \to \photon\particle{H^0} \to \photon\qantiq$.  For
$\unit[20]{GeV} < M_\particle{H^0} < \unit[80]{GeV}$, they find
$\crosssec{\epem \to \particle{H^0} \photon}\, \branching{\particle{H^0}
  \to \qantiq} < \unit[1]{pb}$.  Using Equation~\ref{eqn:pzpwidth} we
translate their full $M_\particle{H^0}$-dependent limits into upper
bounds on $\NTC\anompzp$.  Assuming $\branching{\pngb \to \qantiq}
\approx 1$ and $f_\pngb=\unit[123]{GeV}$, we find limits $\NTC\anompzp
< 1-3$ for PNGB masses below \unit[60]{GeV}, and $\NTC\anompzp < 15$
for PNGB masses below \unit[80]{GeV}.

As Figure~\ref{fig:LEPI:combined-pjj} illustrates, the several limits on
$\NTC\anompzp$ for hadronically-decaying PNGB are similar.  They
improve on the bounds in Reference~\cite{gr-ehs} by a factor of two to
three.

\subsection{Limits on $\NTC\anompzz$}
\label{sec:pzzlimits}

We next obtain limits on \NTC\anompzz\ from the \LEPI\ data.  The
relevant decay paths we examine include $\Znaught \to \Zstar \pngb \to
\missingE\qqbar$ (where the \pngb\ can decay either hadronically or
invisibly) and $\Znaught \to \Zstar \pngb \to \photon\photon\qqbar$,
so that final states with two jets will dominate
(Figure~\ref{fig:fds:lepi:PaZs}).  

  In principle, we must also consider the contribution of an off-shell
  photon to the \qqbar\ production processes
  (Figure~\ref{fig:fds:lepi:PaZs}), which would give a limit on
  \anompzp; however, these results are numerically much weaker than
  the equivalent limits we obtained in Section~\ref{sec:pzplimits}.
  Therefore, we shall apply these limits on $\NTC\anompzz$ only to models
  where $\anompzp \ll \anompzz$, such as the Appelquist-Terning one-family
  model \cite{ta-jt} discussed in Section~\ref{sec:onefamily}.

\subsubsection{$\Znaught \to \Zstar \pngb \to \jet\; \jet \missingE$}
\label{sec:pzz:jjmissing}

\begin{figure}
  \begin{center}
    \includegraphics[width=(\textwidth-1in)/2]{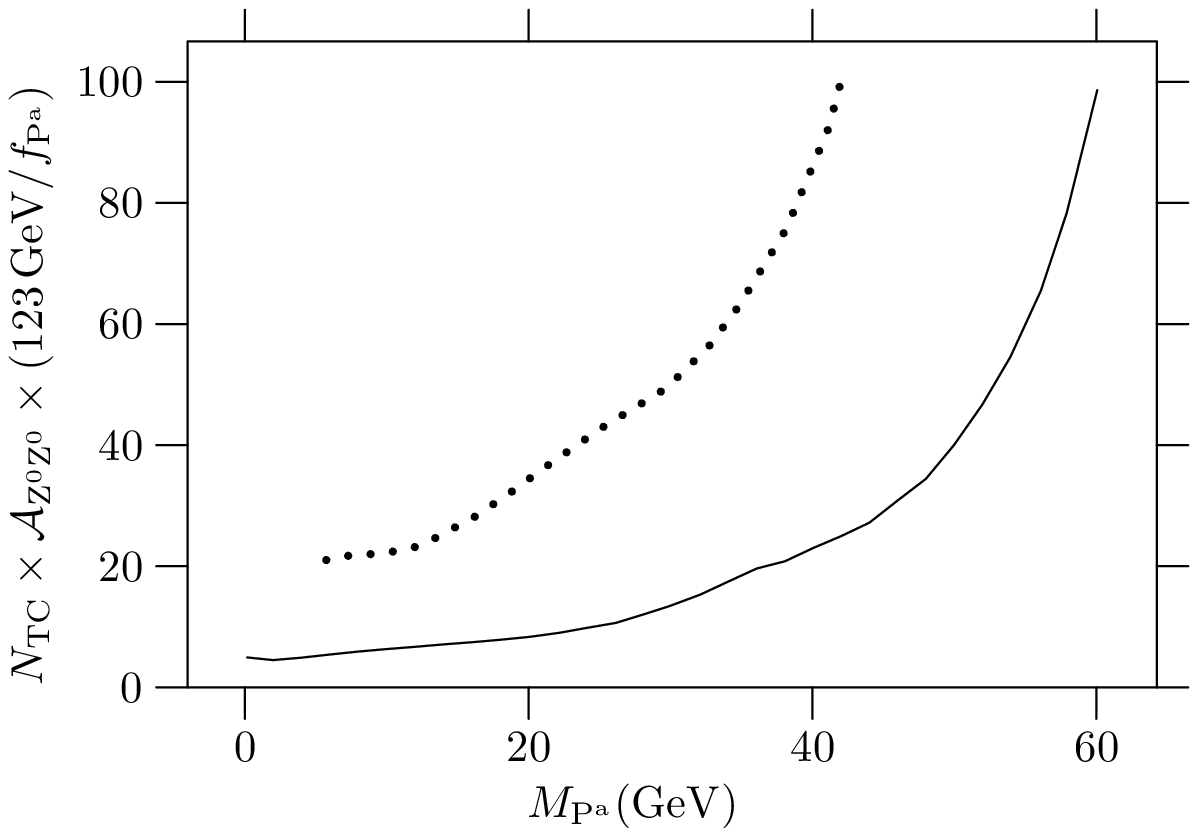}
    \caption{\small Upper limits at 95\% c.l.\ on $\NTC\anompzz\ffrat$ from $\Znaught \to
      \Zstar\pngb \to \qantiq\missingE$, based on \OPAL\ 
      data~\cite{OPALzstarp}. The dotted curve denotes the limits on a
      hadronically decaying \pngb, while the solid curve holds for an
      invisibly decaying \pngb.}
    \label{fig:LEPI:OPALzstarp}
  \end{center}
\end{figure}

This final state can arise in two ways: with the off-shell \Znaught\ 
decaying hadronically and the PNGB decaying invisibly, or with the
off-shell \Znaught\ decaying invisibly (to neutrino pairs) and the
PNGB decaying hadronically (Figure~\ref{fig:fds:lepi:PaZs}).  The
\OPAL\ collaboration has searched for production of a scalar particle,
\particle{S^0}, in both modes, based on $\unit[160]{pb^{-1}}$ of data
collected near the \Znaught\ pole~\cite{OPALzstarp}. They find no
evidence for either mode, and place 95\% c.l.\ upper limits on
the production cross section for $\qantiq\missingE$ through the
intermediate state, $\Zstar\particle{S^0}$, normalized to the
production cross section for
the Standard Model (SM) Higgs $\Zstar\Hzero$ intermediate state
\footnote{The SM Higgs branching ratio can be found in the
  literature~\cite{hhref,hhguide}
  \begin{multline}
    \frac{\branching{\Znaught\to\Hzero\ffbar}}{\branching{ 
        \Znaught \to \ffbar}} = \frac{g^2}{192 \pi^2 
      \cossqtw} \Biggl[ \frac{3y(y^4-8y^2+20)}{\sqrt{4 -
        y^2}}\cos^{-1}\left(\frac{y(3-y^2)}{2}\right) - \nonumber\\
    3(y^4-6y^2+4)\ln 
    y - \frac{1}{2}(1-y^2)(2y^4-13y^2 + 47)\Biggr]\ ,\nonumber
  \end{multline}
  where $y = M_\Hzero/M_\Znaught > \Gamma_\Znaught/M_\Znaught$. This
  approximation neglects the masses of the fermions \particle{f}, and
  the \Znaught\ width, $\Gamma_\Znaught$, which is acceptable for
  $y>\Gamma_\Znaught/M_\Znaught$.  Using this branching ratio, we can
  derive the necessary cross section.}
, $\crosssec{\epem \to \particle{H^0_{\text{SM}}} \Zstar}$.  We call their
ratio of cross sections $R$.  For the visible decay of the scalar, the
numerator of $R$ is $\crosssec{\epem \to \particle{S^0} \Zstar}
\branching{\particle{S^0} \to \qantiq}$, and we label the ratio
$R_{\text{visible}}$.  For $M_\particle{S^0} = \unit[5]{GeV}$, the upper
limit on $R_{\text{visible}}$ is $10^{-3}$; this weakens to
$R_{\text{visible}} \leq 1$ as $M_\particle{S^0}$ increases to
\unit[65]{GeV}.  For the invisible decay of the scalar, the numerator of
$R$ is taken to be $\crosssec{\epem \to \particle{S^0} \Zstar}$, and we label
the ratio $R_{\text{invisible}}$.  The upper limit on
$R_{\text{invisible}}$ is $10^{-4}$ at $M_\particle{S^0} = \unit[0]{GeV}$;
this weakens to $R_{\text{visible}}\leq 1$ as $M_\particle{S^0}$ rises 
toward $M_\Znaught$.

Using Equation~\ref{eqn:pzzwidth}, we derive upper bounds on
$\NTC\anompzz$.  For a PNGB that (nearly) always decays to \qantiq\ 
with $f_\pngb=\unit[123]{GeV}$, we find $\NTC\anompzz < 20-50$ for
PNGB masses below \unit[30]{GeV}.  For an invisibly decaying PNGB, we
find $\NTC\anompzz < 5-13$ for PNGB masses below \unit[30]{GeV}.  In
both cases, above \unit[30]{GeV}, the data become rapidly less
constraining.  Our results appear in Figure~\ref{fig:LEPI:OPALzstarp} .

\subsubsection{$\Znaught \to \Zstar \pngb \to \jet\;\jet\;\photon\photon$}
\label{sec:pzz:jjpp}

\begin{figure}
  \begin{center}
    \includegraphics[width=(\textwidth-1in)/2]{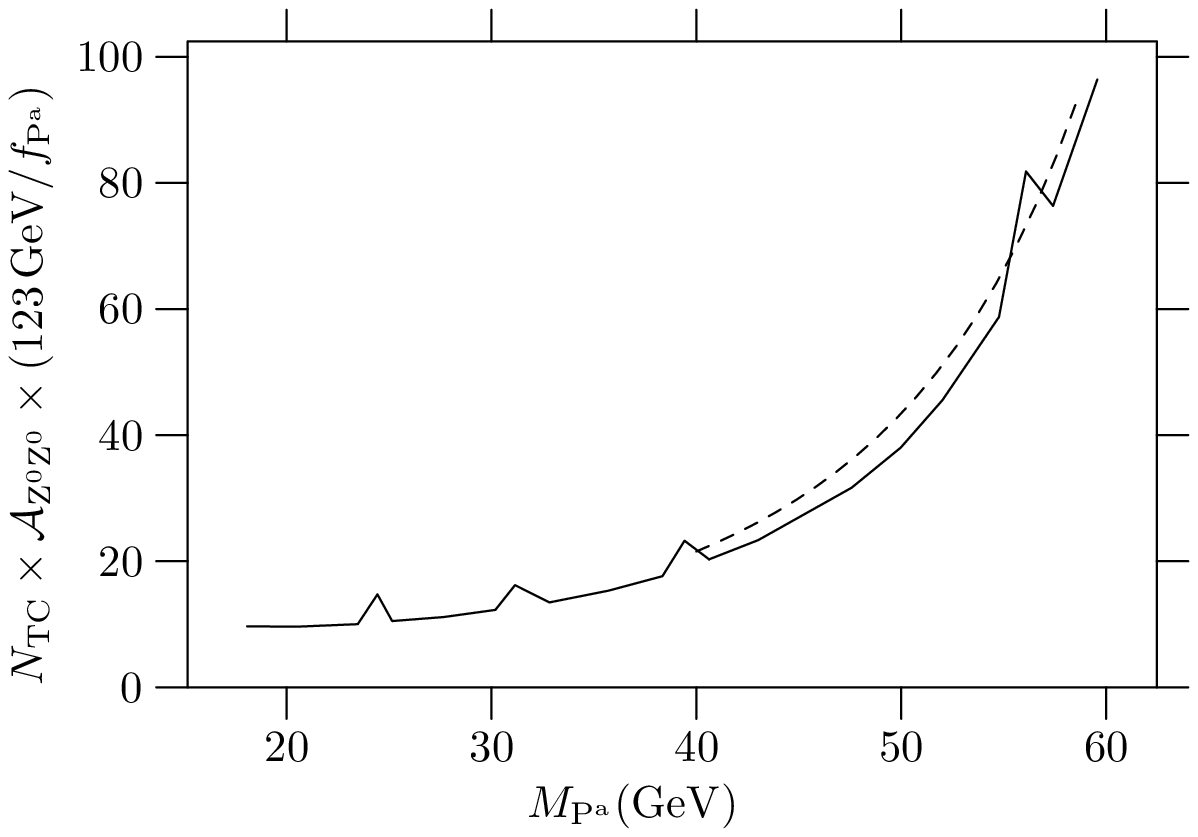}
    \caption{\small Upper limits at 95\% c.l.\ on 
      $\NTC\anompzz\ffrat$ from the process $\Znaught \to \Zstar\pngb \to
      \qantiq\photon\photon$.  Our limits are based on
      \Lthree~\cite{L3hadron} and \OPAL~\cite{OPALhadron} data.  The
      solid curve comes from \Lthree\ data, while the dashed curve
      comes from the \OPAL\ data.  Fluctuations in the curves arise
      from fluctuations in the data.}
    \label{fig:LEPI:combined-jjpp}
  \end{center}
\end{figure}

If the PNGB decays predominantly to photons, a final state with two
hard photons and two jets results (Figure~\ref{fig:fds:lepi:PaZs} with
$\pngb\to \photon\photon$ and $\Zstar\to \qqbar$).  Both the \Lthree\ 
and \OPAL\ collaborations have studied this final state.

\Lthree\ has published limits on the production of a
scalar particle, \particle{H^0}, decaying to two photons and
accompanied by hadrons, based on $\unit[96.8]{pb^{-1}}$ of data
collected near the \Znaught\ pole~\cite{L3hadron}. They find no
evidence for this mode, and place 95\% c.l.\ upper limits on the
production cross section as a function of $M_{\particle{H^0}}$.  For
$\unit[20]{GeV} < M_\particle{H^0} < \unit[70]{GeV}$, the
collaboration finds  $\crosssec{\epem \to \particle{H^0} + \text{
  hadrons}}\, \branching{\particle{H^0} \to \photon\photon} <
\unit[10^{-1}]{pb}$.

The \OPAL\ collaboration has also published limits on the production
of a photonically decaying scalar, \particle{S^0}, in this mode, based
on $\unit[140]{pb^{-1}}$ of data collected on and near the \Znaught\ 
pole~\cite{OPALhadron}. They find no evidence for this mode.  For
particle masses in the range $\unit[40]{GeV} < M_\particle{S^0} <
\unit[80]{GeV}$, \OPAL\ finds a 95\% c.l.\ limit on the product
of branching ratios, $\branching{\Znaught \to \particle{S^0}
  \qantiq}\, \branching{\particle{S^0} \to \photon\photon} <
2\times10^{-6}$.  For smaller masses, $M_\particle{S^0} <
\unit[40]{GeV}$, \OPAL\ states that the limits are weaker, but does
not provide numerical values.

Using Equation~\ref{eqn:pzzwidth}, we infer upper bounds on
$\NTC\anompzz$ in models with PNGB decays dominated by two photon
states and $f_\pngb=\unit[123]{GeV}$.  For PNGB masses below
\unit[30]{GeV}, we find limits $\NTC\anompzz < 10-12$ from the
\Lthree\ results.  In the higher mass range where the \Lthree\ and
\OPAL\ data overlap, they provide nearly identical upper limits on
$\NTC\anompzz$ which become weaker with increasing \pngb\ mass, as
shown in Figure~\ref{fig:LEPI:combined-jjpp}.

\section{Limits from \LEPII}
\label{sec:lepii}
  
In this section we explore the limits that can be obtained on the
anomaly factors \anomppp, \anompzp, and \anompzz\ from published
\LEPII\ data collected at energies well above the \Znaught\ 
pole~\cite{OPAL-LEPII, DELPHI-LEPII, L3-LEPII, DELPHI-missing-LEPII,
  ALEPH-hz-LEPII, L3-hz-LEPII}.  We do so for a number of possible
decay modes of the \pngb; the \pngb\ decay products will be
accompanied either by a hard photon, the decay products of an on-shell
\Znaught, or an \epem\ pair.  In all of the cases that we analyze
below, the final state can arise through either an $s$-channel virtual
photon or \Znaught, or a $2\to 3$ body process.  The Feynman diagrams
for these processes are displayed in Figure~\ref{fig:fds:lepii}. From
Equation~\ref{eqn:epemPaG}, we see that all final states will thus
provide a simultaneous limit, either on \anompzp\ and \anomppp\ (for a
final state photon), or on \anompzz\ and \anompzp\ (for a final state,
on-shell \Znaught).  In all cases, in order to separate these effects,
we first note that the interference term in Equation~\ref{eqn:epemPaG}
is negligible.  In addition, we assume that one or the other of the
direct terms dominates; that assumption is valid in all of the
explicit models we examine in Section~\ref{sec:implications}.  From
Equation~\ref{eqn:epemPaG}, we define the cross sections for processes
with a final state photon by
\begin{equation}
\crossppp = \frac{\alphaem^3 (s - M_\pngb^2)^3}{6 \pi^2 f_\pngb^2 s^3} 
(\NTC \anomppp)^2\ ,
\label{eqn:crossppp}
\end{equation}
for photon-dominated intermediate states, and
\begin{equation}
\crosszpp = \frac{\alphaem^3 (1 - 4\swsq + 8\swfour) (s -
  M_\pngb^2)^3}{ 48 \pi^2 f_\pngb^2 \swfour \cwfour s (s -
  M_\Znaught^2)^2} (\NTC \anompzp)^2\ , 
\label{eqn:crosszpp}
\end{equation}
for \Znaught-dominated intermediate states.  We similarly define the
cross sections for processes with a final state \Znaught\ by
\begin{equation}
\crosszpz = \frac{\alphaem^3 \lambda(s, M_\Znaught^2,
  M_\pngb^2)^{3/2}}{6 \pi^2 f_\pngb^2 \swsq \cwsq s^3} (\NTC
\anompzp)^2\ , 
\label{eqn:crosszpz}
\end{equation}
for photon-dominated intermediate states, and
\begin{equation}
\crosszzz = \frac{\alphaem^3 \lambda(s, M_\Znaught^2,
  M_\pngb^2)^{3/2}}{48 \pi^2 f_\pngb^2 \swsix \cwsix s (s -
  M_\Znaught^2)^2} (\NTC \anompzz)^2\ , 
\label{eqn:crosszzz}
\end{equation}
for \Znaught-dominated intermediate states.  The function $\lambda(a,b,c)$
is given in Section~\ref{sec:equations}.  

In this approximation, the limits set on $\anomppp$ and $\anompzp$ by
processes with a final-state photon are related, because the same data
is being used to separately constrain $\crossppp$ and $\crosszpp$.
The limits that processes with a final-state \Znaught\ set on
$\anompzp$ and $\anompzz$ are, likewise, related.  By comparing the
sizes of the factors preceding $(\NTC\anom)^2$ in equations 4.1 and
4.2 (4.3 and 4.4), one may see that a LEP II limit on $\NTC\anompzp$
($\NTC\anompzz$) is always stronger than the related limit on
$\NTC\anomppp$ ($\NTC\anompzp$), for any PNGB mass.  In any specific
model where the values of the anomaly factors are known, we can
recombine\footnote{For example, take a process with a final-state
  photon and write the theoretical cross section as
\begin{equation*}
\sigma = \Fppp (\NTC\anomppp)^2 + \Fpzp (\NTC\anompzp)^2\ .
\end{equation*}
Then, if the experimental limit $\sigma \le \sigma_{\text{data}}$ is
taken to imply that 
$\NTC^\gamma = \sqrt{\sigma_{\text{data}} / \Fppp\anomppp^2}$ when 
photon-exchange dominates, a little manipulation shows that
the more general limit is
\begin{equation*}
\NTC \leq \NTC^\gamma \left( 1 +
    \left(\frac{\anompzp}{\anomppp}\right)^2 \frac{\Fpzp}{\Fppp}
    \right)^{-\frac{1}{2}}\ .
\end{equation*}.
} the pair of implied limits on \NTC\ from $\crossppp$ and $\crosszpp$
(or from $\crosszpz$ and $\crosszzz$) to obtain a single limit on
\NTC.  We will not need to do this in the models discussed in
Section~\ref{sec:implications}, as one of the paired anomaly factors
always dominates.

\subsection{Processes constraining both $\NTC\anomppp$ and $\NTC\anompzp$}
\label{sec:lepii:appp-apzp}

\subsubsection{Limits from $\epem \to \photon\pngb \to
  \photon\photon\photon$} 
\label{sec:lepii:ppp}

\begin{figure}
\begin{center}
\includegraphics[width=(\textwidth-1in)/2]{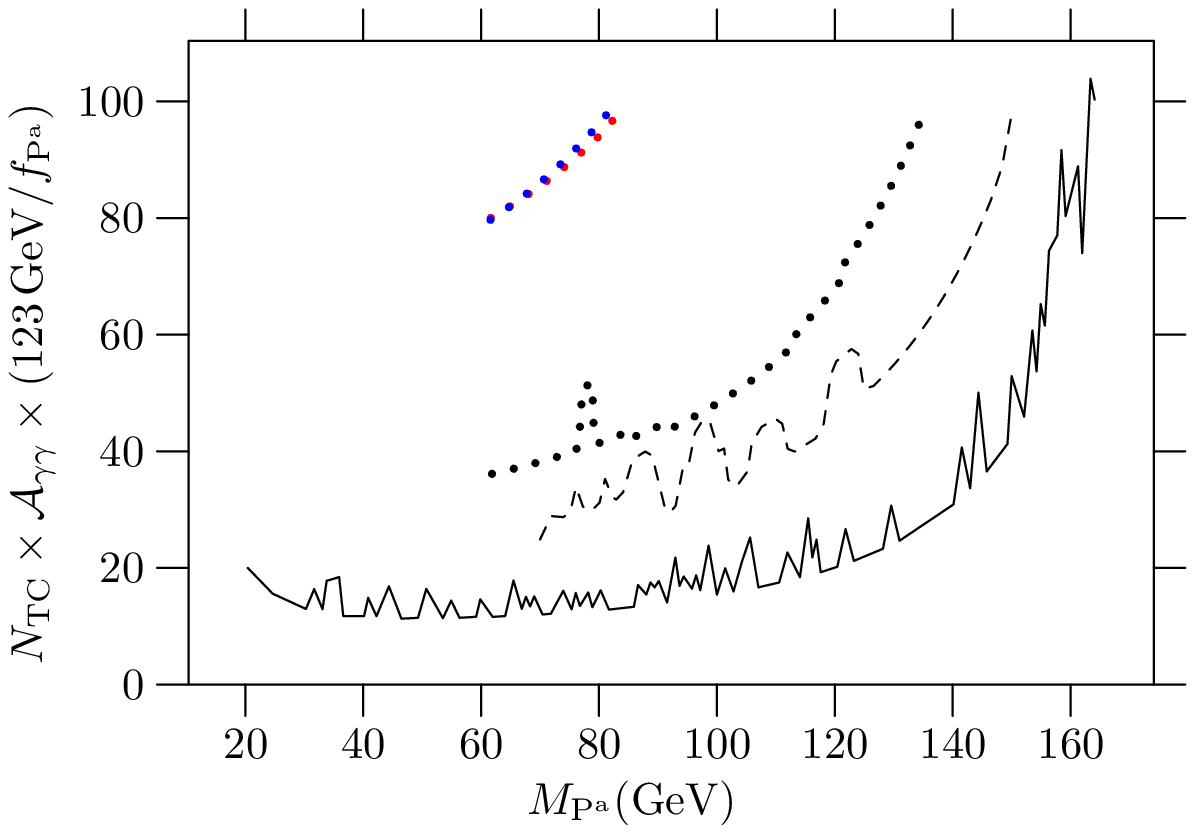}
\qquad
\includegraphics[width=(\textwidth-1in)/2]{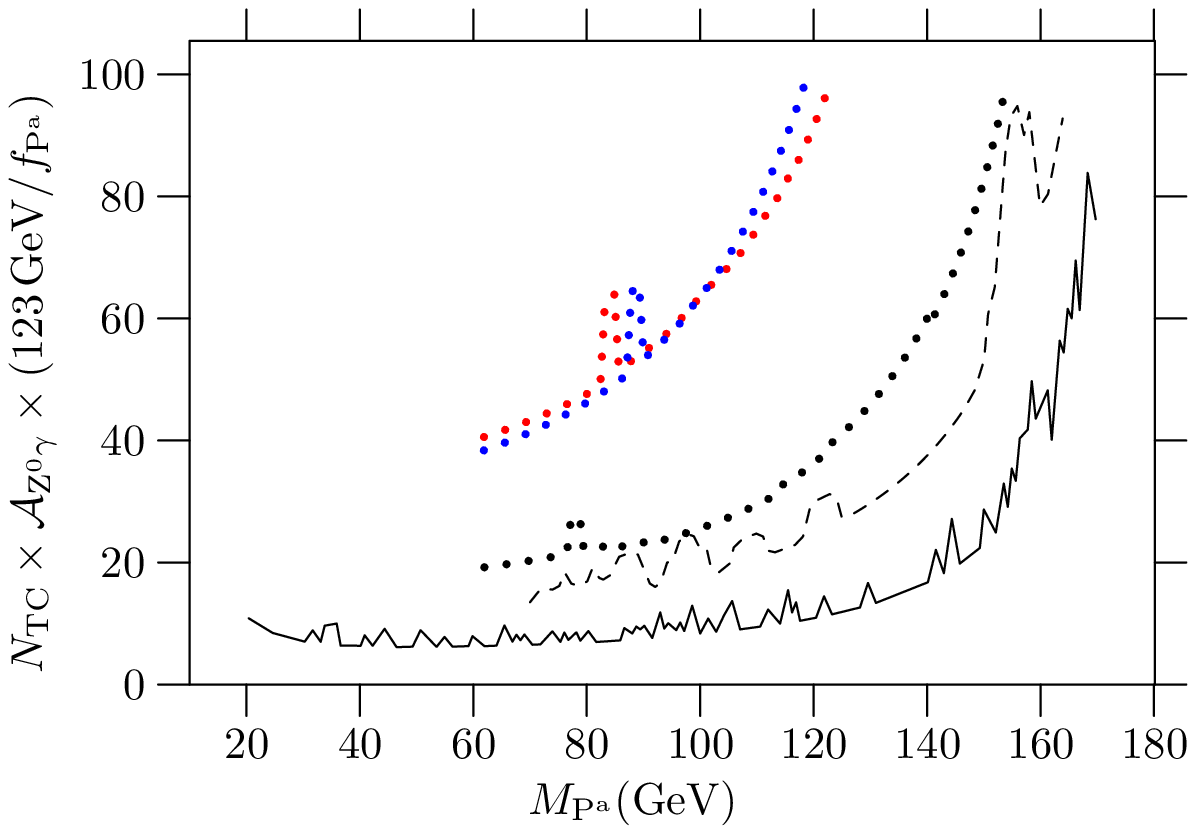}
\caption{Upper limits at 95\% c.l.\ on $\NTC\anomppp\ffrat$ (at left) and $\NTC\anompzp\ffrat$
  (at right) from $\epem \to \photon\pngb \to \photon\photon\photon$.
  The solid line is derived from \OPAL\ data~\cite{OPAL-LEPII}; the
  dashed line is derived from \Lthree\ data~\cite{DELPHI-LEPII}; the
  three dotted lines come from \DELPHI\ data~\cite{DELPHI-LEPII} at
  various center of mass energies (from top to bottom,
  \unit[161]{GeV}, \unit[172]{GeV}, and \unit[183]{GeV}).
  Fluctuations in the curves arise from fluctuations in the data.}
\label{fig:lepii:ppp}
\end{center}
\end{figure}

If the PNGB decays predominantly to photons, the final state can contain
three hard photons (Figure~\ref{fig:fds:lepii:PaP} with $\pngb\to
\photon\photon$).  The \DELPHI, \Lthree, and \OPAL\ collaborations have all
published limits on this final state from their \LEPII\ data samples.

The \DELPHI\ collaboration has published limits on the production of a scalar
resonance, \particle{H}, decaying to photons.  They have performed three
analyses, based on $\unit[9.7]{pb^{-1}}$ of data collected at $\sqrt{s} =
\unit[161]{GeV}$, $\unit[10.1]{pb^{-1}}$ of data collected at $\sqrt{s} =
\unit[172]{GeV}$, and $\unit[47.7]{pb^{-1}}$ of data collected at $\sqrt{s} =
\unit[183]{GeV}$ \cite{DELPHI-LEPII}. They find no evidence for a new
resonance, and place 95\% c.l.\ upper limits on the cross section
$\crosssec{\epem \to \particle{H}\photon} \branching{\particle{H} \to
  \photon\photon}$, as a function of $M_\particle{H}$.  From data taken at
$\sqrt{s} = \unit[183]{GeV}$, they find $\crosssec{\epem \to
  \particle{H}\photon}\ \branching{\particle{H} \to \photon\photon} <
\unit[0.20]{pb}$ within the mass range $\unit[60]{GeV} < M_\particle{H} <
\unit[184]{GeV}$, almost independent of $M_\particle{H}$.  The data taken 
at lower energies is less constraining (see Figure~\ref{fig:lepii:ppp}).

The \Lthree\ collaboration has published limits on the production of a
scalar resonance, \particle{H}, decaying to photons.  They have
performed an analysis based on $\unit[176]{pb^{-1}}$ of data collected
at $\sqrt{s} = \unit[189]{GeV}$ \cite{L3-LEPII}. They find no evidence
for a new resonance, and place 95\% c.l.\ upper limits on the
cross section $\crosssec{\epem \to \particle{H}\photon}
\branching{\particle{H} \to \photon\photon}$, as a function of
$M_\particle{H}$.  For $\unit[70]{GeV} < M_\particle{H} <
\unit[170]{GeV}$, they find $\crosssec{\epem \to \particle{H}\photon}\ 
\branching{\particle{H} \to \photon\photon} < \unit[0.30]{pb}$, almost 
independent of $M_\particle{H}$.

The \OPAL\ collaboration has also published limits on the production
of a resonance, \particle{X}, decaying to photons.  They have
performed an analysis based on $\unit[178]{pb^{-1}}$ of data collected
at $\sqrt{s} = \unit[189]{GeV}$ \cite{OPAL-LEPII}. They find no
evidence for a new resonance, and place 95\% c.l.\ upper limits
on the cross section $\crosssec{\epem \to \particle{X}\photon}
\branching{\particle{X} \to \photon\photon}$, as a function of
$M_\particle{X}$.  For $\unit[50]{GeV} < M_\particle{X} <
\unit[150]{GeV}$, they find $\crosssec{\epem \to \particle{X}\photon}
 \branching{\particle{X} \to \photon\photon} < \unit[0.03]{pb}$,
roughly independent of $M_\particle{X}$; for masses on either end of
this range, their cross section limit becomes rapidly less
constraining.  This limit is almost an order of magnitude
stronger than either the \Lthree\ or \DELPHI\ limits on the same
process. 

The most stringent limits come from the OPAL data.  Using
Equations~\ref{eqn:crossppp} and~\ref{eqn:crosszpp}, we translate these data
into upper bounds on $\NTC\anomppp$ and $\NTC\anompzp$.  Assuming that \pngb\ 
decay to photons dominates, $\branching{\pngb \to \photon\photon} \approx 1$
and $f_\pngb = \unit[123]{GeV}$, we find $\NTC\anomppp < 15$ for $M_\pngb <
M_\Znaught$; for $M_\pngb < \unit[140]{GeV}$, we find that $\NTC\anomppp <
40$.  We find $\NTC\anompzp < 9$ for $M_\pngb < M_\Znaught$; for $M_\pngb <
\unit[140]{GeV}$, we find that $\NTC\anompzp < 17$.  For larger masses, both
limits become rapidly less constraining.  We plot our results based on the
data from all three collaborations in Figure~\ref{fig:lepii:ppp}.

\subsubsection{Limits from $\epem \to \photon\pngb \to \photon\bantib$}
\label{sec:lepii:pbb}

\begin{figure}
\begin{center}
\includegraphics[width=(\textwidth-1in)/2]{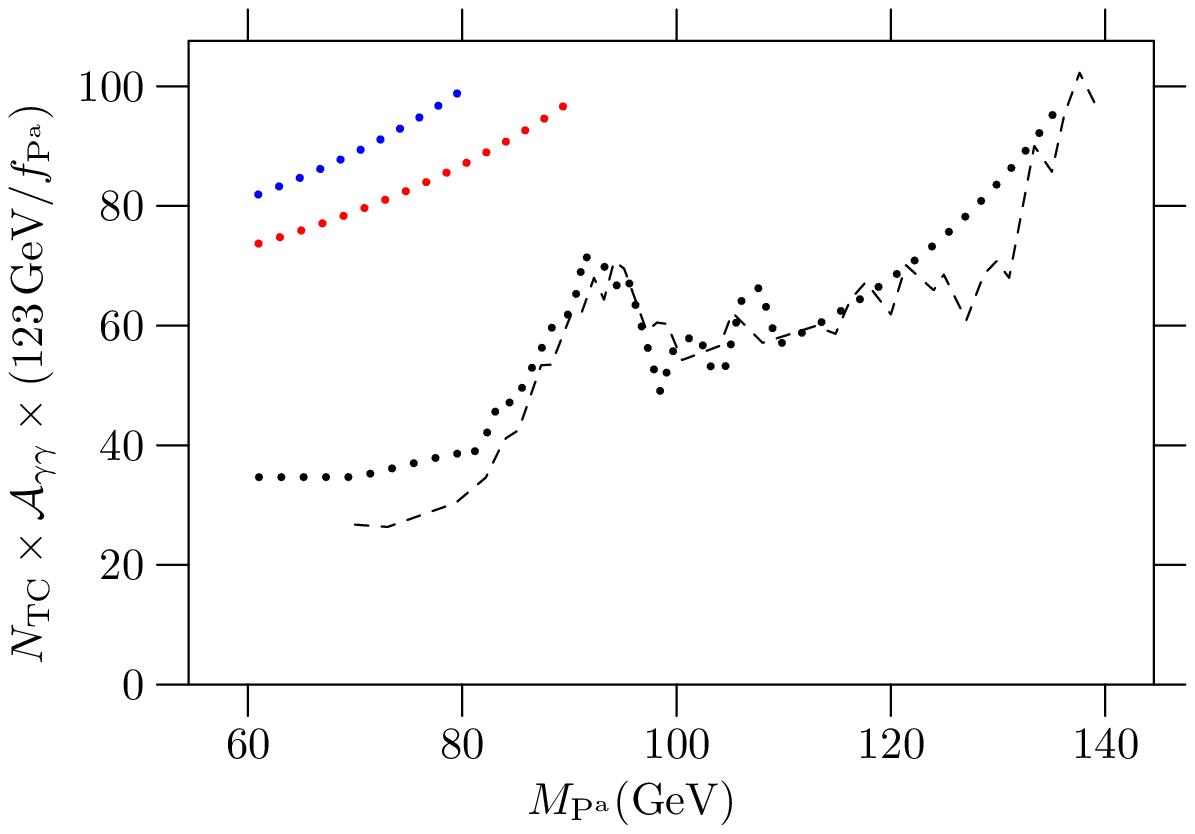}
\qquad
\includegraphics[width=(\textwidth-1in)/2]{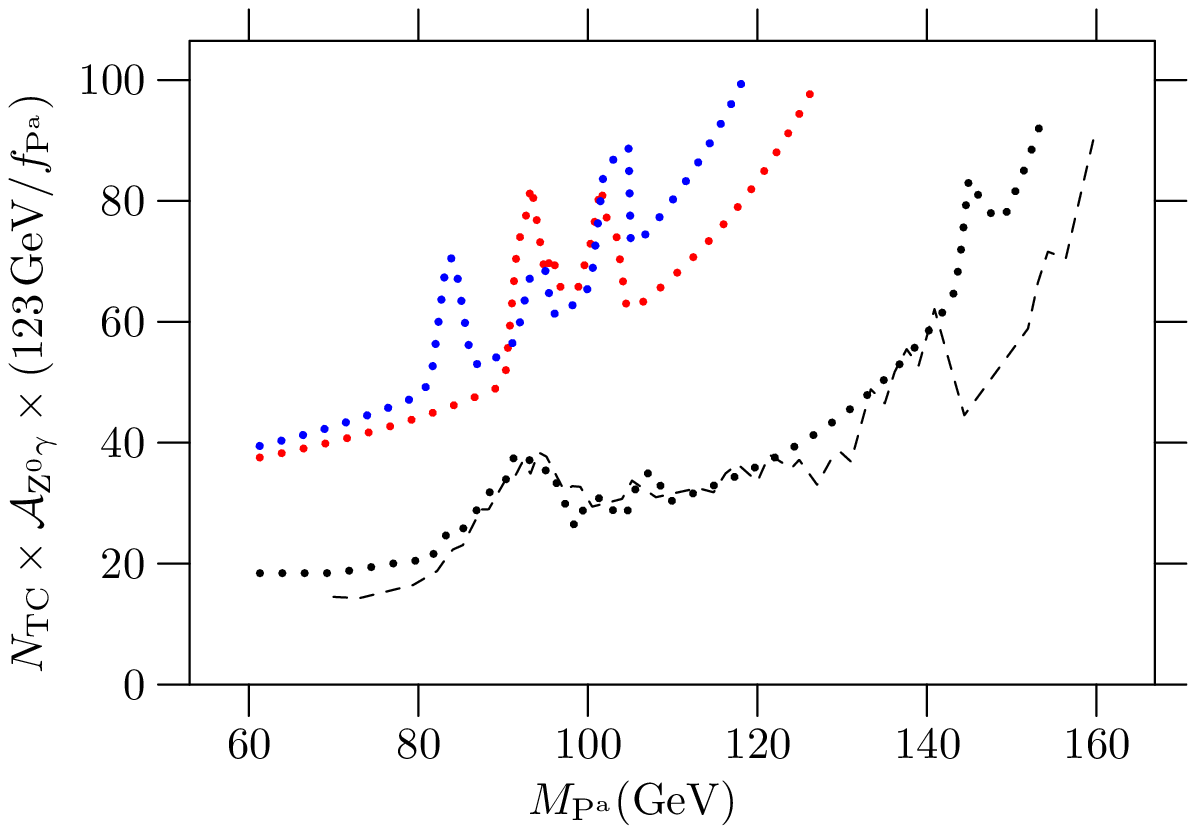}
\caption{Upper limits at 95\% c.l.\ on $\NTC\anomppp\ffrat$ (at left) and $\NTC\anompzp\ffrat$
  (at right) from $\epem \to \photon\pngb \to \photon\bbbar$.  The
  dashed line is derived from \Lthree\ data~\cite{L3-LEPII}; the three
  dotted lines are derived from \DELPHI\ data at various center of
  mass energies (from top to bottom, \unit[161]{GeV}, \unit[172]{GeV},
  and \unit[183]{GeV}).  Fluctuations in the curves arise from
  fluctuations in the data.}
\label{fig:lepii:pbb}
\end{center}
\end{figure}

If the PNGB decays predominantly to \bbbar\ pairs, the final state can
contain a hard photon, and two \qbottom\ jets
(Figure~\ref{fig:fds:lepii:PaP} with $\pngb\to \bbbar$).  The \DELPHI\ 
and \Lthree\ collaborations have both published limits from their
\LEPII\ data samples.

The \DELPHI\ collaboration has published limits on the production of a
scalar resonance, \particle{H}, decaying to \bbbar\ pairs.  They have
performed this analysis at each of three center of mass energies,
based on $\unit[9.7]{pb^{-1}}$ of data collected at $\sqrt{s} =
\unit[161]{GeV}$ , $\unit[10.1]{pb^{-1}}$ of data collected at
$\sqrt{s} = \unit[172]{GeV}$, and $\unit[47.7]{pb^{-1}}$ of data
collected at $\sqrt{s} = \unit[183]{GeV}$ \cite{DELPHI-LEPII}. They
find no evidence for a new resonance, and place 95\% c.l.\ upper
limits on the cross section $\crosssec{\epem \to \particle{H}\photon}
\branching{\particle{H} \to \bbbar}$, as a function of
$M_\particle{H}$.  Their highest-energy data is the most constraining;
for $\unit[60]{GeV} < M_\particle{H} < \unit[184]{GeV}$, they find
$\crosssec{\epem \to \particle{H}\photon} \branching{\particle{H} \to
  \bbbar} < \unit[0.50]{pb}$, almost independent of $M_\particle{H}$.

The \Lthree\ collaboration has published limits on the production of a
scalar resonance, \particle{H}, decaying to \bbbar\ pairs.  They have
performed this analysis on $\unit[176]{pb^{-1}}$ of data collected at
$\sqrt{s} = \unit[189]{GeV}$ \cite{L3-LEPII}. They find no evidence for
a new resonance, and place 95\% c.l.\ upper limits on the cross
section $\crosssec{\epem \to \particle{H}\photon}
\branching{\particle{H} \to \bbbar}$, as a function
of $M_\particle{H}$.  For $\unit[70]{GeV} < M_\particle{H} <
\unit[170]{GeV}$, they find $\crosssec{\epem \to \particle{H}\photon}
\branching{\particle{H} \to \bbbar} < \unit[0.30]{pb}$.

Using Equations~\ref{eqn:crossppp} and~\ref{eqn:crosszpp}, we can
translate this data into upper bounds on both $\NTC\anomppp$ and
$\NTC\anompzp$.  Assuming the \pngb\ decays predominantly to \bbbar\ 
jets, $\branching{\pngb \to \bbbar} \approx 1$ and $f_\pngb
=\unit[123]{GeV}$, we find that $\NTC\anomppp < 62$ for $M_\pngb <
M_\Znaught$; for $M_\pngb < \unit[140]{GeV}$, we find that
$\NTC\anomppp < 140$.  We find $\NTC\anompzp < 30$ for $M_\pngb <
M_\Znaught$; for $M_\pngb < \unit[140]{GeV}$, we find that
$\NTC\anompzp < 60$.  For larger masses, both limits become rapidly
less constraining.  We plot our results based on the data from both
collaborations in Figure~\ref{fig:lepii:pbb}.

\subsubsection{Limits from $\epem \to \photon\pngb \to \photon\missingE$}
\label{sec:lepii:pmissing}

\begin{figure}
\begin{center}
\includegraphics[width=(\textwidth-1in)/2]{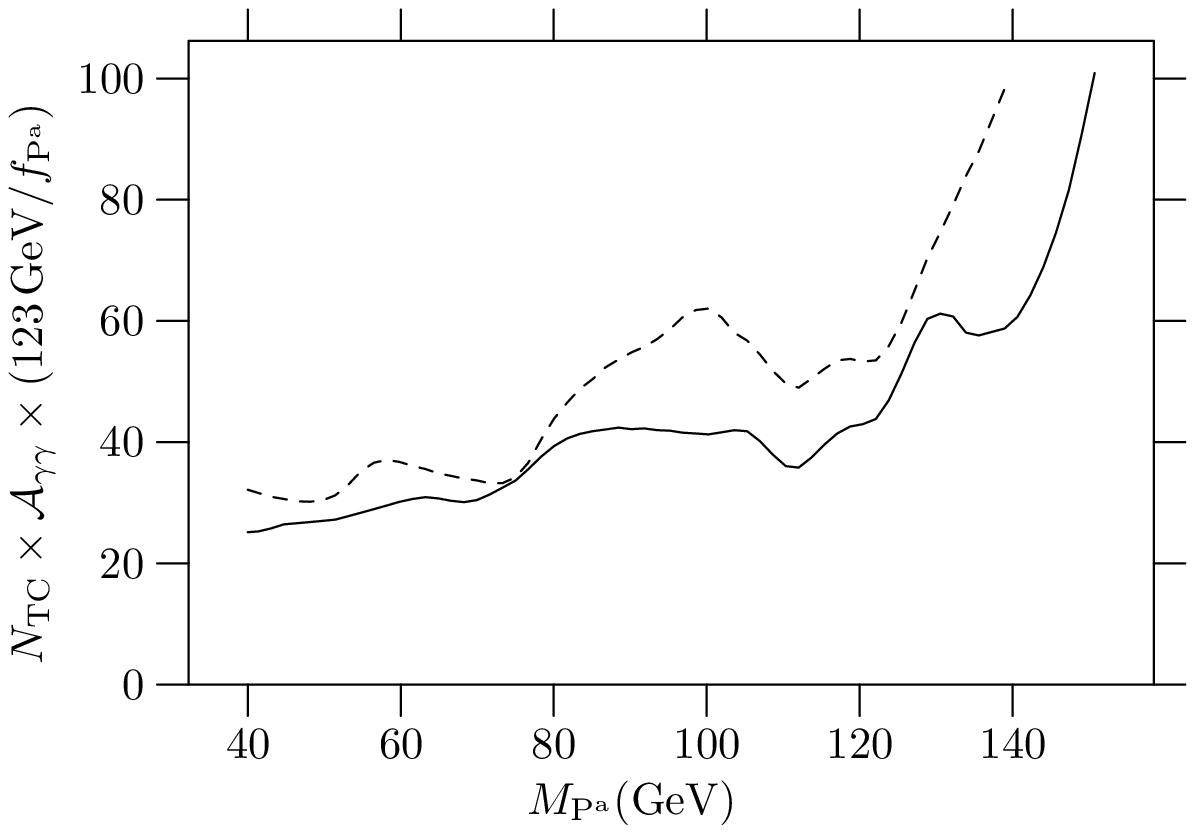}
\qquad
\includegraphics[width=(\textwidth-1in)/2]{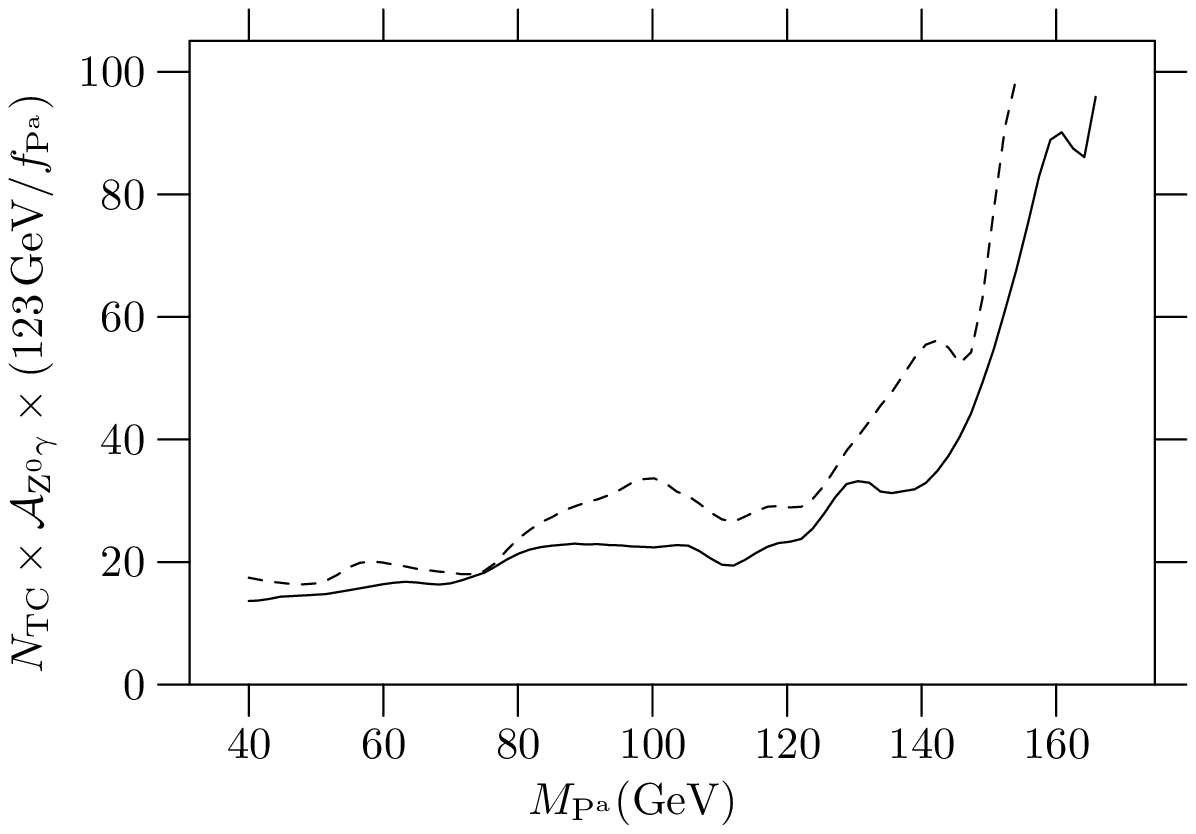}
\caption{Upper limits at 95\% c.l.\ on $\NTC\anomppp\ffrat$ (at left) and $\NTC\anompzp\ffrat$
  (at right) from $\epem \to \photon\pngb \to \photon\missingE$.  The
  results were derived from \DELPHI\ data~\cite{DELPHI-missing-LEPII};
  the solid lines come from the stronger limit derived by \DELPHI,
  while the dashed lines correspond to the weaker limit.  Fluctuations
  in the curves arise from fluctuations in the data.}
\label{fig:lepii:pmissing}
\end{center}
\end{figure}

If the predominant decays of the PNGB are invisible, we can find
at \LEPII\ a final state with a single hard photon and missing energy
(Figure~\ref{fig:fds:lepii:PaP} with $\pngb\to \missingE$).  The
\DELPHI\ collaboration has searched for anomalous single photon events
produced by a new scalar particle, \particle{X}, in
$\unit[51]{pb^{-1}}$ of data collected at $\unit[183]{GeV}$ and in
$\unit[158]{pb^{-1}}$ collected at
$\unit[189]{GeV}$ \cite{DELPHI-missing-LEPII}. They find no evidence
for a new resonance, and place 95\% c.l.\ upper limits on the
production cross section $\sigma_\particle{X}$ as a function of
$M_\particle{X}$.  They provide two limits, based on their inclusion
of data from different calorimeters: for $\unit[40]{GeV} <
M_\particle{X} < \unit[160]{GeV}$, the stronger (weaker) limit is $\sigma_\particle{X} <
\unit[0.2]{pb}\ (\unit[0.3]{pb})$.

Using Equations~\ref{eqn:crossppp} and~\ref{eqn:crosszpp}, we
translate these data into upper bounds on $\NTC\anomppp$ and
$\NTC\anompzp$.  Assuming that invisible decays of the \pngb\ 
dominate, $\branching{\pngb \to \missingE} \approx 1$ and $f_\pngb =
\unit[123]{GeV}$, we find $\NTC\anomppp < 40$ for $M_\pngb <
M_\Znaught$; for $M_\pngb < \unit[140]{GeV}$, we find $\NTC\anomppp <
60$.  We find $\NTC\anompzp < 23$ for $M_\pngb < M_\Znaught$; for
$M_\pngb < \unit[140]{GeV}$, we find $\NTC\anompzp < 33$. We plot our
results in Figure~\ref{fig:lepii:pmissing}.

\subsection{Processes constraining both $\NTC\anompzp$ and $\NTC\anompzz$}
\label{sec:lepii:apzp-apzz}

In order to place limits on $\anompzz$ from \LEPII\ data, we need to
find states which include both intermediate and final \Znaught\ bosons
coupled to the \pngb.  Unfortunately, the most general processes that
include these states also include three other diagrams, which receive
contributions not only from \anompzz, but also from \anomppp\ and
\anompzp (Figure~\ref{fig:fds:lepii:PaPsZs}).\footnote{This is not an
  issue for SM Higgs searches, since there are no tree level couplings
  of the Higgs to photons.}

In this section, we explore a restricted set of processes, those which
include a real \Znaught\ in the final state
(Figure~\ref{fig:fds:lepii:PaZ}).  In the context of experiments, this
involves requiring that the final state visible energy which is 
assumed not to come from the \pngb\ satisfies an invariant mass
constraint, $M_{\text{visible}} \approx M_\Znaught$.  While this
simplifies the analysis significantly, it reduces both the number of
available published analyses, and the range of PNGB masses that are
accessible, such that $M_\pngb < \sqrt{s} - M_\Znaught$.  The \LEPII\ 
data collected at $\sqrt{s} = \unit[189]{GeV}$ for example, can only
probe PNGB masses lighter than about $\unit[95]{GeV}$.

\subsubsection{Limits from $\epem \to \pngb (\photon^*/\Zstar) \to
  \ffbar\missingE$}
\label{sec:lepii:ppffbar}

\begin{figure}
\begin{center}
\includegraphics[width=(\textwidth-1in)/2]{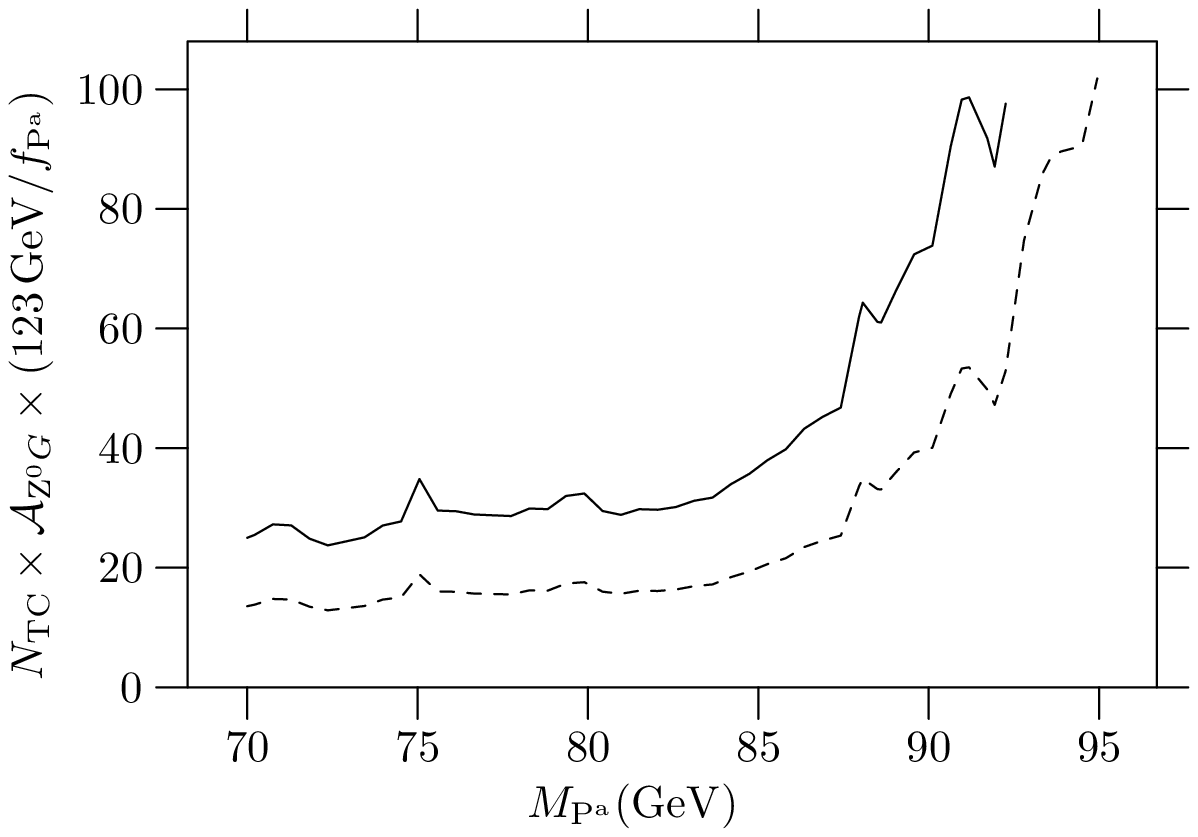}
\caption{Upper limits at 95\% c.l.\ on $\NTC\anompzp\ffrat$ (solid line) and $\NTC\anompzz\ffrat$
  (dashed line) from $\epem \to \Znaught\pngb \to \ffbar\missingE$, where
  the PNGB decays invisibly.  The results were derived from \ALEPH\ 
  data~\cite{ALEPH-hz-LEPII}.  Fluctuations in the curves arise from
  fluctuations in the data.}
\label{fig:lepii:hz-missing}
\end{center}
\end{figure}

If the PNGB is produced in association with a real \Znaught, the final
state can contain missing energy from the PNGB decay, and two fermions
from the \Znaught\ decay (Figure~\ref{fig:fds:lepii:PaZ}).  The
\ALEPH\ collaboration has searched in this mode for the production of
a scalar boson, \higgs, in $\unit[172]{pb^{-1}}$ of data collected at
$\unit[189]{GeV}$ \cite{ALEPH-hz-LEPII}. To insure that the visible
energy comes from a \Znaught, the collaboration requires that the
invariant mass of the visible decay products approximately equal the
invariant mass of the \Znaught, $M_\ffbar \approx M_\Znaught$.  They
find no evidence for a new resonance, and place 95\% c.l.\ upper
limits on the cross section for $\higgs\Znaught$ production,
scaled to the SM cross section,
\footnote{The SM $\epem \to \Higgs\Znaught$ cross
section can be found in the literature~\cite{hhguide,hzref1,hzref2}  
\begin{equation}
\crosssec{\epem \to \Higgs\Znaught} = \frac{\pi \alphaem^2
  (1 +(1-4\swsq)^2)}{192 \swfour \cwfour s^2 (s - M_\Znaught^2)^2}
\Lambda^{1/2} (\Lambda + 12 s M_\Znaught^2)\ ,
\end{equation}
where $\Lambda = \lambda(s, M_\Znaught^2, M_\Higgs^2)$, as defined in
Section~\ref{sec:equations}. 
 }
via $\branching{\higgs \to \missingE} \crosssec{\epem \to
  \higgs\Znaught}/ \crosssec{\epem \to \higgs\Znaught}_{{\text{SM}}}$,
which we label $R$.  For $M_\pngb < \unit[85]{GeV}$, the upper limit
is approximately $R < 0.1$; for larger \pngb\ masses, the limit rises
rapidly to $R < 1$ at $M_\pngb = \unit[95]{GeV}$.

Using Equations~\ref{eqn:crosszpz} and~\ref{eqn:crosszzz}, we translate these
data into upper bounds on $\NTC\anompzp$ and $\NTC\anompzz$.  Assuming the
\pngb\ predominantly decays into invisible states and that
$f_\pngb=\unit[123]{GeV}$, we find that $\NTC\anompzp < 20$ for $M_\pngb <
\unit[85]{GeV}$, with the limit rapidly weakening for larger masses.  For
$M_\pngb < \unit[85]{GeV}$, we find that $\NTC\anompzz < 30$.  We plot our
results in Figure~\ref{fig:lepii:hz-missing}

\subsubsection{Limits from $\epem \to \pngb (\photon^*/\Zstar) \to
  \ffbar\photon\photon$}  
\label{sec:lepii:mEffbar}

\begin{figure}
\begin{center}
\includegraphics[width=(\textwidth-1in)/2]{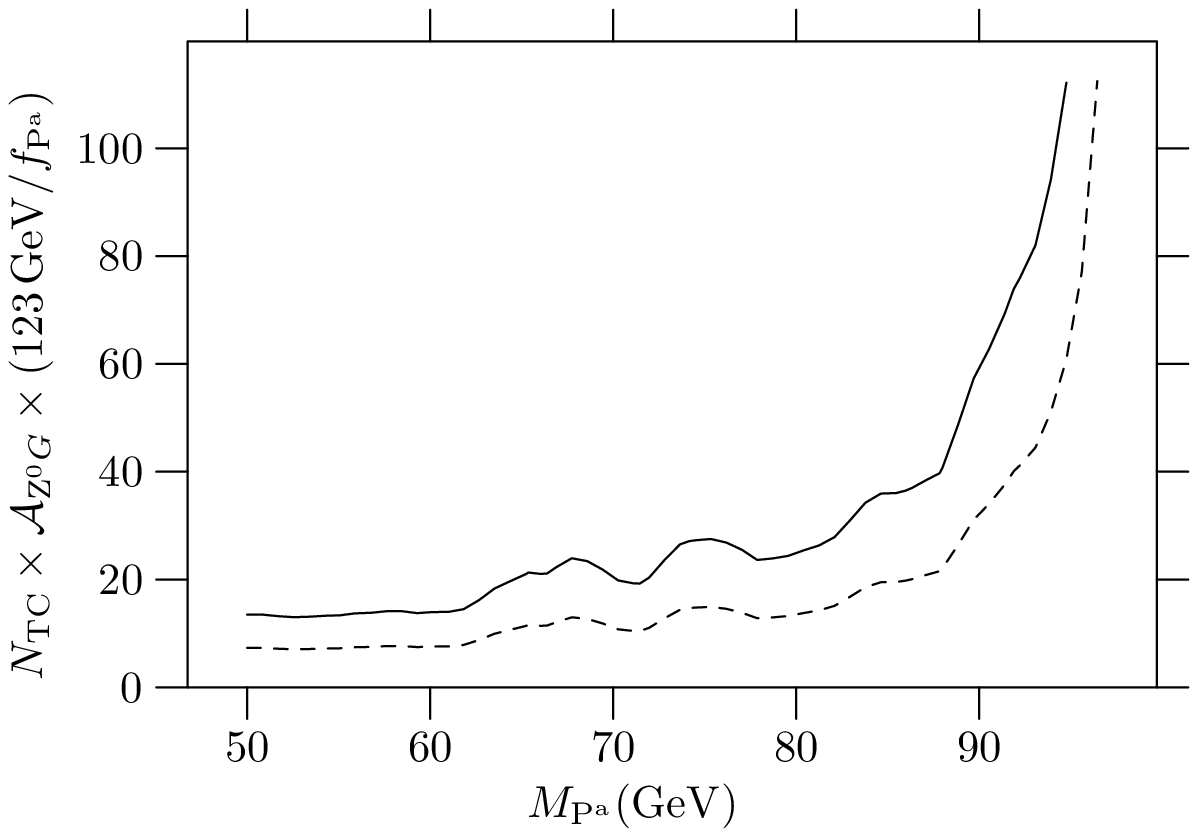}
\caption{Upper limits at 95\% c.l.\ on $\NTC\anompzp\ffrat$ (solid line) and $\NTC\anompzz\ffrat$
  (dashed line) from $\epem \to \Znaught\pngb \to \ffbar\photon\photon$.
  The results were derived from \Lthree\ data~\cite{L3-hz-LEPII}.
  Fluctuations in the curves arise from fluctuations in the data.}
\label{fig:lepii:hz-pp}
\end{center}
\end{figure}

If the PNGB is produced in association with a real \Znaught, the final
state can contain two photons from the PNGB decay, and two fermions
from the \Znaught\ decay (Figure~\ref{fig:fds:lepii:PaZ}).  The
\Lthree\ collaboration has searched in this mode for the production of
a scalar boson, \higgs, in $\unit[176]{pb^{-1}}$ of data collected at
$\unit[189]{GeV}$ \cite{L3-hz-LEPII}. The collaboration requires that
the fermions come from a real \Znaught\ by applying an invariant mass
cut.  They find no evidence for a new resonance, and place 95\% c.l.\ 
upper limits on the cross section for $\higgs\Znaught$ production,
scaled to the SM cross section (given in the previous section), via $R
= \branching{\higgs \to \photon\photon} \crosssec{\epem \to
  \higgs\Znaught}/ \crosssec{\epem \to \higgs\Znaught}_{{\text{SM}}}$.
For $M_\pngb < \unit[85]{GeV}$, the upper limit is approximately $R <
0.1$; for larger masses, the limit rises rapidly to $R < 1$ at
$M_\pngb = \unit[98]{GeV}$.

Using Equation~\ref{eqn:crosszpz} and~\ref{eqn:crosszzz}, we translate these
data into upper bounds on $\NTC\anompzp$ and $\NTC\anompzz$.  Assuming the
\pngb\ predominantly decays into photon pairs and $f_\pngb=\unit[123]{GeV}$,
we find that $\NTC\anompzp < 15$ for $M_\pngb < \unit[85]{GeV}$, with the
limit rapidly weakening for larger masses.  For $M_\pngb < \unit[85]{GeV}$,
we find that $\NTC\anompzz < 25$.  We plot our results in
Figure~\ref{fig:lepii:hz-pp}.

\subsection{Process $\epem \to \pngb\epem$ constraining $\NTC\anomppp$}
\label{sec:lepii:appp}

\begin{figure}
\begin{center}
\includegraphics[width=(\textwidth-1in)/2]{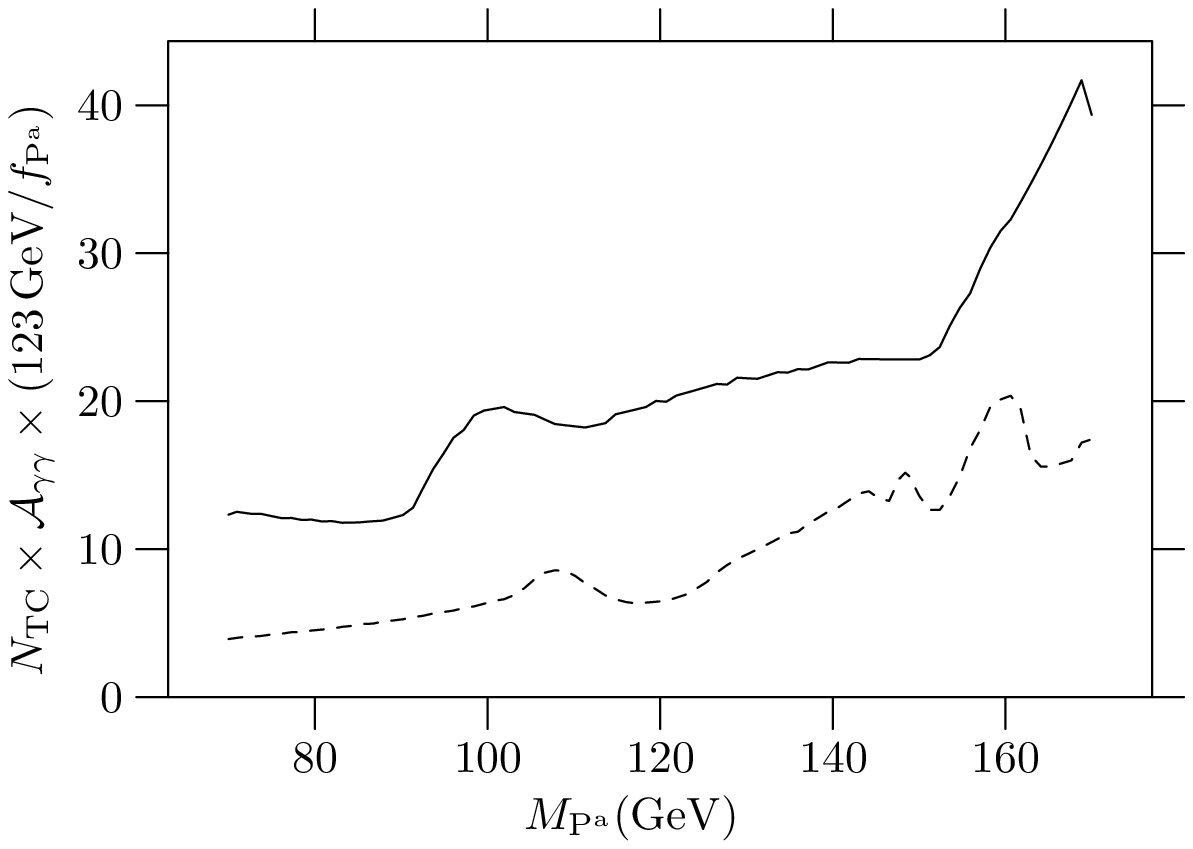}
\caption{Upper limits at 95\% c.l.\ on $\NTC\anomppp\ffrat$ from $\epem \to \pngb\epem$,
  derived from \Lthree\ data.  The solid curve holds if
  $\branching{\pngb \to \bbbar} \approx 1$, while the dashed curve
  holds if $\branching{\pngb \to \photon\photon} \approx 1$.
  Fluctuations in the curves arise from fluctuations in the data. }
\label{fig:lepii:width-limits}
\end{center}
\end{figure}

Since the \pngb\ couples to the electroweak gauge bosons, it is
possible to produce them in the $2 \to 3$ interaction, $\epem \to
\pngb\epem$ (Figure~\ref{fig:fds:lepii:Paepem}).  The \Lthree\ 
collaboration has performed a search for anomalous couplings of a SM
Higgs boson, \Higgs, to electroweak gauge bosons in
$\unit[176]{pb^{-1}}$ of data collected at
$\unit[189]{GeV}$ \cite{L3-LEPII}. They find no evidence for such
anomalous couplings, and place 95\% c.l.\ upper limits on the
decay widths $\Gamma_\bbbar = \width{\Higgs \to \photon\photon}
\branching{\Higgs \to \bbbar}$ and $\Gamma_{\photon\photon} =
\width{\Higgs \to \photon\photon} \branching{\Higgs \to
  \photon\photon}$, as a function of $M_\Higgs$.  They find
$\Gamma_{\photon\photon} < \unit[10^{-1}]{MeV}$ for $M_\Higgs <
\unit[70]{GeV}$, rising to $\Gamma_{\photon\photon} <
\unit[10^2]{GeV}$ at $M_\Higgs < \unit[170]{GeV}$; the limits on
$\Gamma_\bbbar$ are approximately an order of magnitude larger at all
$M_\Higgs$.

Using Equation~\ref{eqn:PaPPwidth}, we translate these data into upper 
bounds on\footnote{The upper bound is only on $\anomppp$, rather than some
  combination of $\anomppp$, $\anompzp$ and $\anompzz$, because kinematic
  factors ensure that the gauge bosons internal to the $2\to 3$ process are
  predominantly photons~\cite{vl-ps, L3-LEPII}.}
  $\NTC\anomppp$.  Assuming the photon decay mode of the
\pngb\ dominates and $f_\pngb=\unit[123]{GeV}$, we 
find $\NTC\anomppp < 5$ for $M_\pngb < M_\Znaught$; for $M_\pngb <
\unit[140]{GeV}$, we find $\NTC\anomppp < 10$.  If instead the \pngb\
decays predominantly to \bbbar , we find $\NTC\anomppp < 12$ for $M_\pngb < M_\Znaught$, and
$\NTC\anomppp < 20$ for $M_\pngb < \unit[140]{GeV}$.  We plot our
results in Figure~\ref{fig:lepii:width-limits}.

\section{Summarizing the \LEPI\ and \LEPII\ Limits}
\label{sec:combined}

\begin{figure}
\begin{center}
\subfigure[$\pngb\to \photon\photon$]{\includegraphics[width=
  (\textwidth-1in)/3]{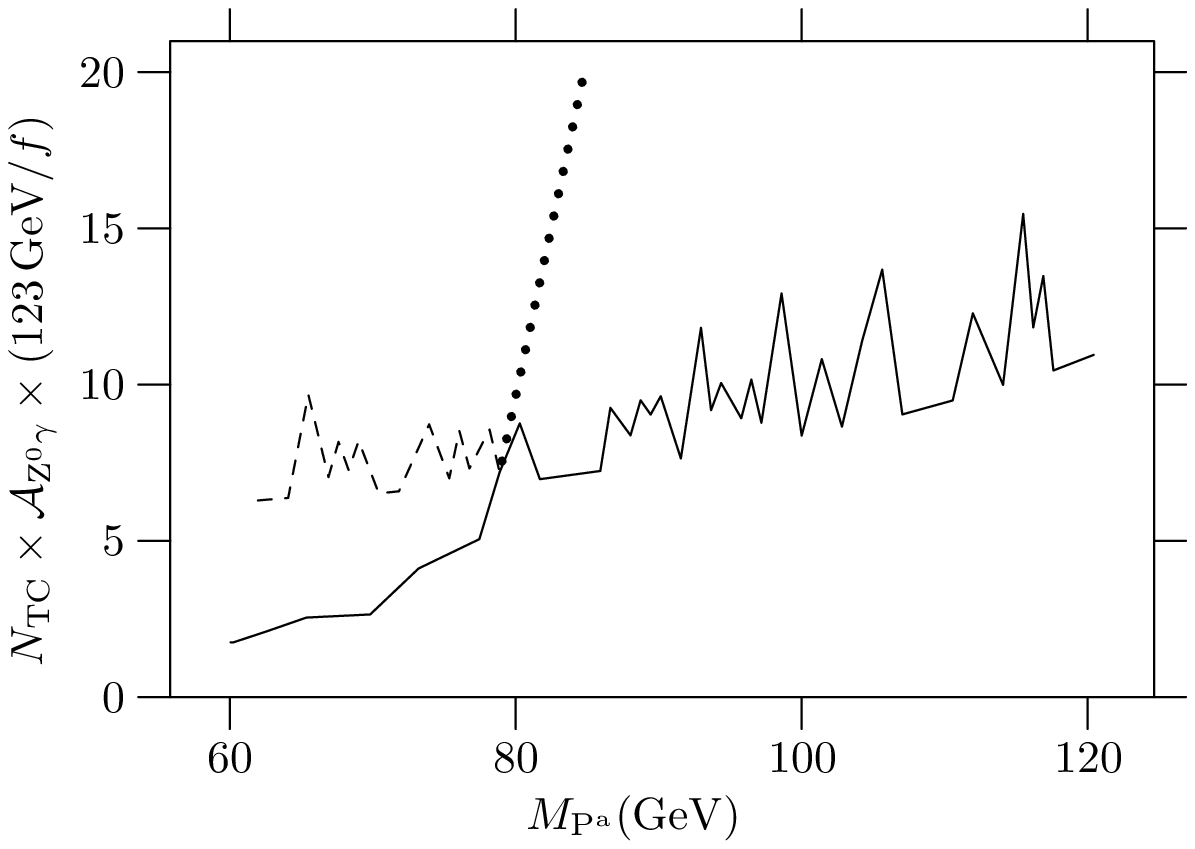} \label{fig:overlap-pzp:PaPP}}
\qquad
\subfigure[$\pngb\to \bbbar$]{\includegraphics[width=
  (\textwidth-1in)/3]{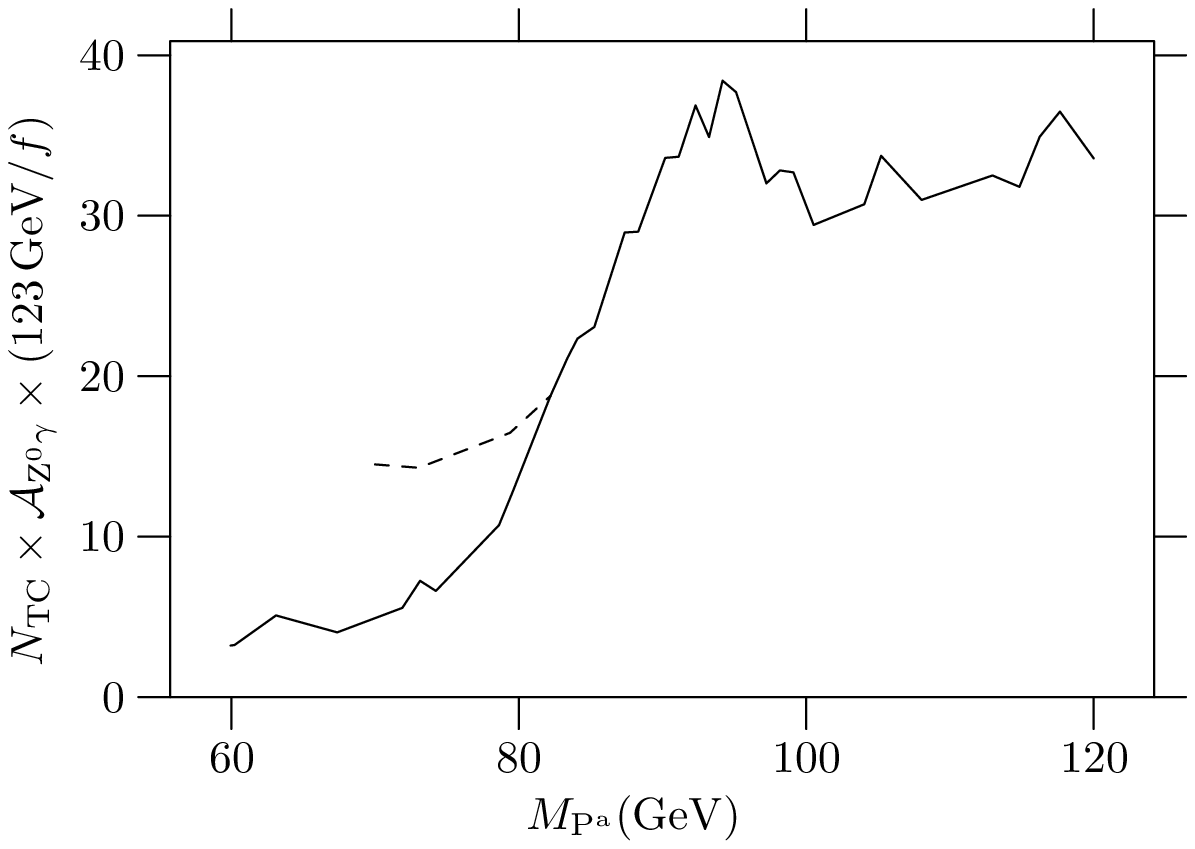} \label{fig:overlap-pzp:Pabb}}
\qquad
\subfigure[$\pngb\to \missingE$]{\includegraphics[width=
  (\textwidth-1in)/3]{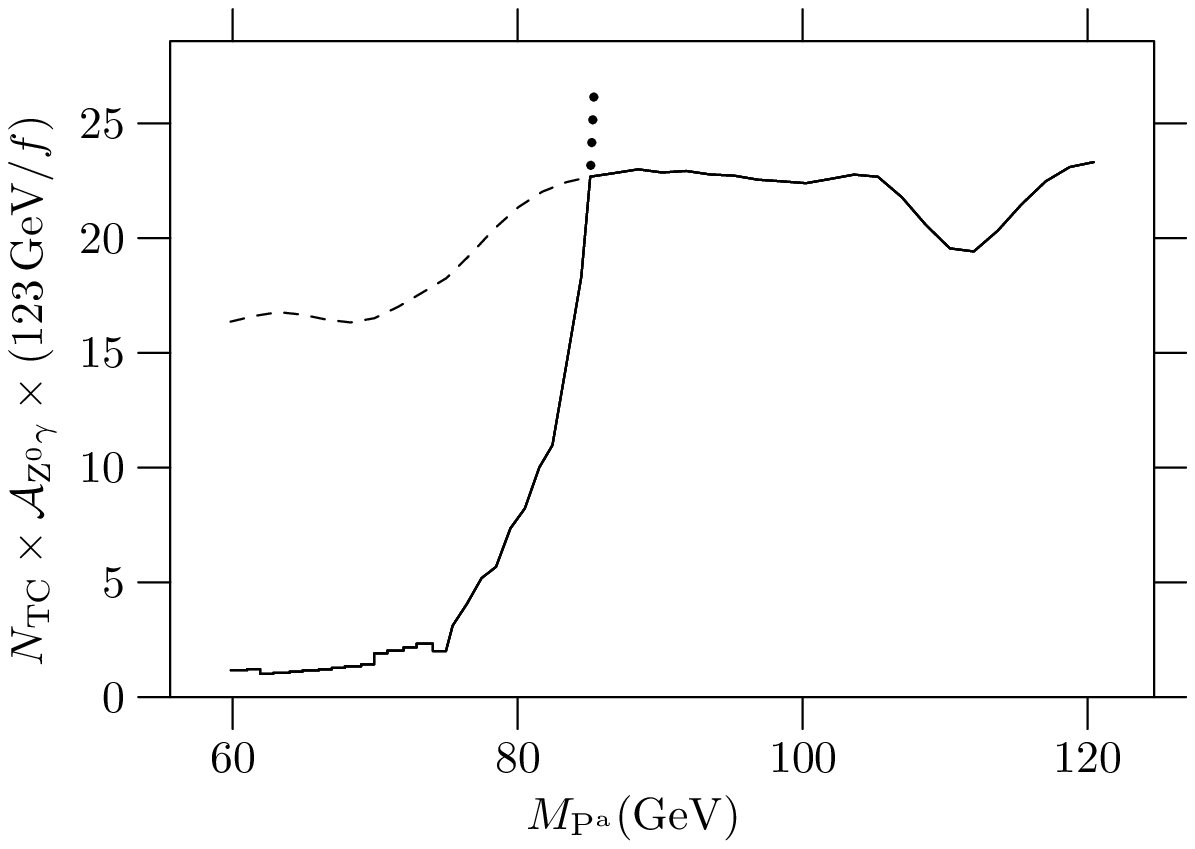} \label{fig:overlap-pzp:PamE}}
\caption{Combined \LEPI\ and \LEPII\ upper limits at 95\% c.l.\ on $\NTC\anompzp\ffrat$
  from $\photon\photon\photon$ (left), $\photon\bantib$ (center), and
  $\photon\missingE$ (right) final states, for $M_\pngb$ within
  \unit[30]{GeV} of $M_\Znaught$.  In each case, the solid line
  indicates the combined limit, the dotted line indicates the \LEPI\
  limit, and the dashed line indicates the \LEPII\ limit.}
\label{fig:overlap-pzp}
\end{center}
\end{figure}

\begin{figure}
\begin{center}
\subfigure[$\pngb\to \photon\photon$]{\includegraphics[width=
  (\textwidth-1in)/3]{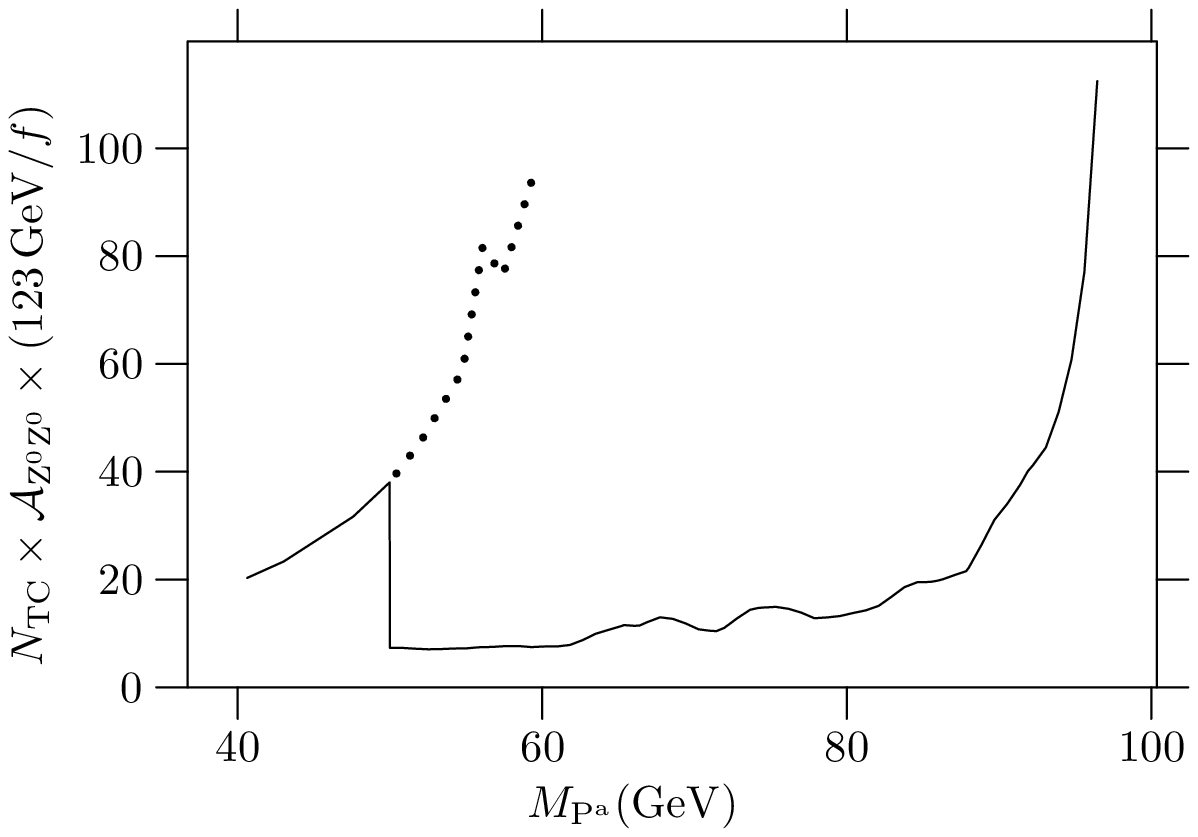} \label{fig:overlap-pzz-PaPP}}
\qquad
\subfigure[$\pngb\to \missingE$]{\includegraphics[width=
  (\textwidth-1in)/3]{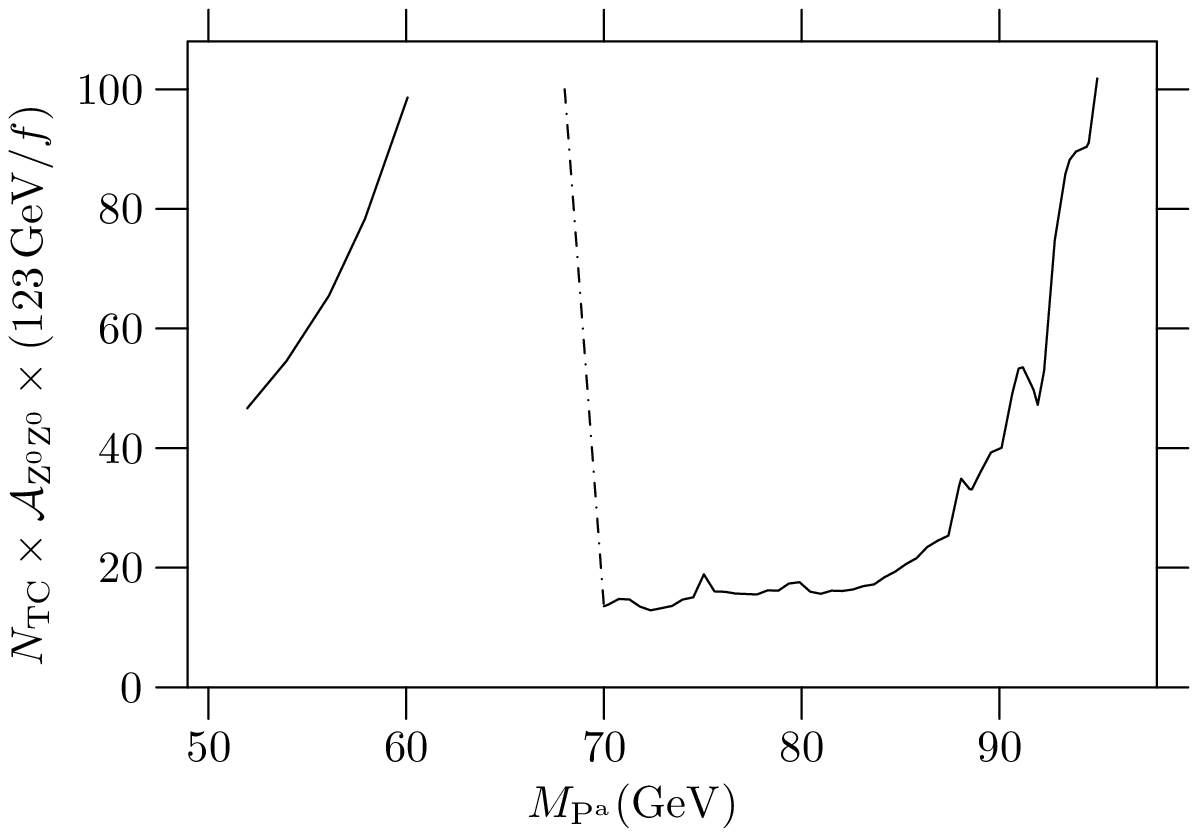} \label{fig:overlap-pzz-PamE}}
\caption{Combined \LEPI\ and \LEPII\ upper limits at 95\% c.l.\ on $\NTC\anompzz\ffrat$
  from $\pngb \to \photon\photon$ (at left) and $\pngb \to \missingE$
  (at right).  In each case, the solid limit indicates the combined
  limit and the dotted line indicates the \LEPI\ limits.  The
  dotted-dashed line in the right hand plot is only to guide the eye
  since the data sets sensitive to low-mass (\LEPI) and high-mass
  (\LEPII) PNGBs do not overlap).}
\label{fig:overlap-pzz}
\end{center}
\end{figure}

\begin{table}
  \caption{Upper limits on $\NTC\anom(\unit[123]{GeV}/f_\pngb)$ from
    the \LEPI\ and \LEPII\ data samples.  The limits in this table are
    independent of each other, and can be applied directly to any
    model.} 
  \label{tab:independent}
\vspace{\baselineskip}
  \begin{center}
    \begin{tabular}{|c|c|c|c|c|c|c|c|c|c|c|c|c|}   \hline
      PNGB  
         & \multicolumn{4}{|c|}{produced via \anomppp}
         & \multicolumn{4}{|c|}{produced via \anompzp}
         & \multicolumn{4}{|c|}{produced via \anompzz}\\ \hline
        mass &\multicolumn{4}{c|}{decay mode}
         &\multicolumn{4}{c|}{decay mode}
         &\multicolumn{4}{c|}{decay mode}\\ \hline
         $M_\pngb \leq$ & 
         $\missingE$ &
         $\photon\photon$ &
         $\bantib$ &
         $\jetj\jetj$ & 
         $\missingE$ &
         $\photon\photon$ &
         $\bantib$ &
         $\jetj\jetj$ &
         $\missingE$ &
         $\photon\photon$ &
         $\bantib$ &
         $\jetj\jetj$ \\ \hline
      $\unit[30]{GeV}$ &&&&& 0.63 & 0.75 & 1.1 & 1.3 & 13 & 12 && 50 \\
      $\unit[60]{GeV}$ &&&&& 1.2 & 1.8 & 3.2 & 3.1 &&&&\\
      $\unit[80]{GeV}$ && 5 & 12 && 7.8 & 10 & 14 & 13 &&&&\\ \hline
      $\unit[100]{GeV}$ && 6 & 19 &&&&&&&&&\\
      $\unit[120]{GeV}$ && 7 & 20 &&&&&&&&&\\
      $\unit[140]{GeV}$ && 13 & 23 &&&&&&&&&\\
      $\unit[160]{GeV}$ && 19 & 32 &&&&&&&&&\\ \hline
    \end{tabular} 
  \end{center}
\end{table}

\begin{table}
  \caption{Upper limits on $\NTC\anom(\unit[123]{GeV}/f_\pngb)$ from
    the \LEPI\ and \LEPII\ data samples.  The limits in this table are 
    not independent: a given final state simultaneously
    provides two limits: for either \anomppp\ and \anompzp\ or
    \anompzp\ and \anompzz.  For a given mass and \pngb\ decay mode,
    the related limits on \anomppp\ and \anompzp\ [\anompzp\ and \anompzz ]
    are surrounded by  parentheses [brackets]; e.g. for an
    \unit[80]{GeV} PNGB
    decaying to two photons, the limits (14) and (8) are related, as are the
    limits [25] and [14]. Each limit in this table is derived under the
    assumption that its production process (anomaly factor) dominates, as
    discussed in~\ref{sec:lepii}.  For models in which the various anomaly
    factors are of quite different sizes (as in all models studied in
    Section~\ref{sec:implications}), the strongest limit from the
    table applies 
    directly.  For a model in which the two related production modes are
    comparable, the limits can be combined as discussed in
    Section~\ref{sec:lepii} to obtain a stronger bound on \NTC.  }   
  \label{tab:dependent}
\vspace{\baselineskip}
  \begin{center}
    \begin{tabular}{|c|c|c|c|c|c|c|c|c|c|c|c|c|}   \hline
      PNGB 
         & \multicolumn{4}{|c|}{produced via \anomppp}
         & \multicolumn{4}{|c|}{produced via \anompzp}
         & \multicolumn{4}{|c|}{produced via \anompzz}\\ \hline
     mass    &\multicolumn{4}{c|}{decay mode}
         &\multicolumn{4}{c|}{decay mode}
         &\multicolumn{4}{c|}{decay mode}\\ \hline
         $M_\pngb \leq$ & 
         $\missingE$ &
         $\photon\photon$ &
         $\bantib$ &
         $\jetj\jetj$ & 
         $\missingE$ &
         $\photon\photon$ &
         $\bantib$ &
         $\jetj\jetj$ &
         $\missingE$ &
         $\photon\photon$ &
         $\bantib$ &
         $\jetj\jetj$ \\ \hline
      $\unit[30]{GeV}$ && (13) &&&& (8) &&&&&&\\
      $\unit[60]{GeV}$ & (30) & (12) &&& (16) & (7)/[14] &&&& [8] &&\\
      $\unit[80]{GeV}$ & (39) & (14) & (30) && (21)/[30] & (8)/[25]
      & (17) && [16] & [14] &&\\ \hline
      $\unit[100]{GeV}$ & (42) & (18) & (58) && (23) & (10) & (32) &&&&& \\
      $\unit[120]{GeV}$ & (44) & (22) & (66) && (23) & (12) & (36) &&&&& \\
      $\unit[140]{GeV}$ & $(60)$ & (36) & (98) && (33) & (20) & (53)
      &&&&& \\ 
      $\unit[160]{GeV}$ & $(165)$ & (82) & $(172)$ && (86) & (45) &
      (100) &&&&& \\ \hline 
    \end{tabular} 
  \end{center}
\end{table}

In this section, we summarize and compare the limits derived from the \LEPI\ 
and \LEPII\ data sets.  First, we graphically examine the region of overlap
between the \LEPI\ and \LEPII\ data sets.  Then, we tabulate our derived
limits on the various anomaly factors.

Both \LEPI\ and \LEPII\ provide access to \anompzp\ and \anompzz.  In
Figure~\ref{fig:overlap-pzp}, we display the region of overlap between the
\anompzp\ results from \LEPI\ and the \anompzp-dominated limits from \LEPII.
We find that for all decay modes (except $\pngb \to \jetj\jetj$, which is not
probed at \LEPII), the \LEPI\ data provide a much stronger limit than the
\LEPII\ data for $M_\pngb < \unit[80]{GeV}$, while for $M_\pngb >
\unit[80]{GeV}$, the \LEPII\ data take over.  Figure~\ref{fig:overlap-pzz}
similarly displays the limited region of overlap between the \anompzz\ 
dominated results from \LEPI\ and the \anompzz\ dominated limits from \LEPII.
Here, the \LEPI\ data exist only up to $M_\pngb \approx \unit[60]{GeV}$ and the
\LEPII\ data are stronger than \LEPI\ data where they exist.

Tables~\ref{tab:independent} and~\ref{tab:dependent} gather the best limits
on $\NTC\anom(\unit[123]{GeV}/f_\pngb)$ from the experiments discussed
in Sections~\ref{sec:lepi} and~\ref{sec:lepii}.  In
Table~\ref{tab:independent}, we gather all limits that can be
independently applied to TC models; that is, these limits are not
directly linked with any other anomaly factor limits.  In
Table~\ref{tab:dependent}, we gather all limits that can not be
independently applied; that is, the limits on \anomppp\ and \anompzz\ in this
table are related to the corresponding limits on \anompzp , as discussed in
Section~\ref{sec:lepii}.  In particular, it is permissible to apply
these limits directly only if the appropriate anomaly factor dominates
the \pngb\ production (as in the models we examine in
Section~\ref{sec:implications}).

\section{Implications for Technicolor Models}
\label{sec:implications}

In this section, we discuss how our limits on \pngb\ couplings
constrain several classes of technicolor models.  We begin with a
quick look at the familiar one-family technicolor models in order to
assess what properties a model must have in order that our limits
constrain the masses of the PNGBs in that model.  We then examine
three other scenarios: near-critical Extended Technicolor models,
models with weak isotriplet technifermions, and low-scale models.
Because the data are sensitive to the ratio $\NTC\anom / f_\pngb$ (per
Equations~\ref{eqn:pzpwidth} and~\ref{eqn:pzzwidth}), models with
smaller technipion decay constants or larger anomaly factors will be
more tightly constrained.

\subsection{One-family Technicolor Models}
\label{sec:onefamily}

\begin{table}
\caption{Upper limits on the number of technicolors, \NTC, as a 
  function of \pngb\ decay constant and PNGB mass in the
  Applequist-Terning one-family technicolor model of
  Section~\ref{sec:onefamily}.  The superscripted 
  labels indicate the data used to calculate the limits: \dag:
  \anomppp\ from Table~\ref{tab:independent}; \ddag: \anompzp\ from
  Table~\ref{tab:independent}; \$: \anompzz\ from
  Table~\ref{tab:independent}; \hashh: \anomppp\ from
  Table~\ref{tab:dependent}.} 
\label{tab:revenge}
\vspace{\baselineskip}
\begin{center}
  \begin{tabular}{|c|c|c|} \hline
    & \multicolumn{2}{|c|}{$\NTC \le$} \\ \hline
    $M_\pngb \le$ & $\PEtechni \to \photon\photon$ & $\PNtechni \to
    \missingE$ \\ \hline
    \unit[30]{GeV} & $37 f_\PEtechni/v$\uphashh & $295 f_\PNtechni/v$\updollar\\ 
    \unit[60]{GeV} & $34 f_\PEtechni/v$\uphashh & --- \\
    \unit[80]{GeV} & $14 f_\PEtechni/v$\updag & $364 f_\PNtechni/v$\updollar\\ 
    \unit[100]{GeV} & $17 f_\PEtechni/v$\updag & --- \\
    \unit[120]{GeV} & $20 f_\PEtechni/v$\updag & --- \\
    \unit[140]{GeV} & $37 f_\PEtechni/v$\updag & --- \\
    \unit[160]{GeV} & $54 f_\PEtechni/v$\updag & --- \\ \hline
  \end{tabular}
\end{center}
\end{table}

\begin{table}
    \caption{Upper limits on the number of technicolors, \NTC, as a
      function of the technifermion hypercharge $y$ and PNGB mass in
      the Manohar-Randall one-family weak-isotriplet TC model~\cite{am-lr} of 
      Section~\ref{sec:onefamily}.  Limits are shown for 
      the cases where the dominant decays are invisible, two-photon,
      or \bbbar.  The superscripted labels indicate the data used to 
      calculate the limits: \dag: \anomppp\ from
      Table~\ref{tab:independent}; \ddag: \anompzp\ from
      Table~\ref{tab:independent}; \hashh: \anomppp\ from
      Table~\ref{tab:dependent}.}
\label{tab:isotriplet}
\vspace{\baselineskip}
  \begin{center}
    \begin{tabular}{|c|c|c|c|c|c|}   \hline
      \multicolumn{2}{|c|}{PNGB} & \multicolumn{3}{c|}{$\NTC\leq$} \\ 
      \hline 
      & \multicolumn{1}{|c}{$M_\pngb \leq$} & 
      \multicolumn{1}{|c}{$\pngb \to \missingE$} &
      \multicolumn{1}{|c}{$\pngb \to \photon\photon$} &
      \multicolumn{1}{|c|}{$\pngb \to \bantib$} \\ \hline
      & $\unit[30]{GeV}$ & $0.16/y^2$\upddag & $0.19/y^2$\upddag &
      $0.28/y^2$\upddag \\ 
      & $\unit[60]{GeV}$ & $0.31/y^2$\upddag & $0.44/y^2$\upddag &
      $0.82/y^2$\upddag \\ 
      & $\unit[80]{GeV}$ & $2.00/y^2$\upddag & $0.29/y^2$\updag &
      $0.71/y^2$\updag \\ 
       $\Pone$ & $\unit[100]{GeV}$ & $2.47/y^2$\uphashh & $0.35/y^2$\updag
& $1.12/y^2$\updag \\ 
      & $\unit[120]{GeV}$ & $2.59/y^2$\uphashh & $0.41/y^2$\updag &
      $1.18/y^2$\updag \\ 
      & $\unit[140]{GeV}$ & $3.54/y^2$\uphashh& $0.77/y^2$\updag & $1.36/y^2$\updag
      \\ 
      & $\unit[160]{GeV}$ & $9.72/y^2$\uphashh & $1.12/y^2$\updag & $1.89/y^2$\updag
      \\ \hline 
      & $\unit[30]{GeV}$ & $2.28/y$\upddag & $0.94/y$\uphashh &
      $3.96/y$\upddag \\ 
      & $\unit[60]{GeV}$ & $2.17/y$\uphashh & $0.87/y$\uphashh &
      $11.5/y$\upddag \\ 
      & $\unit[80]{GeV}$ & $2.81/y$\uphashh & $0.36/y$\updag &
      $0.87/y$\updag \\ 
\Pthree      & $\unit[100]{GeV}$ & $3.03/y$\uphashh &
$0.43/y$\updag & $1.37/y$\updag \\ 
      & $\unit[120]{GeV}$ & $3.18/y$\uphashh & $0.51/y$\updag &
      $1.44/y$\updag \\ 
      & $\unit[140]{GeV}$ & $4.33/y$\uphashh & $0.94/y$\updag & $1.66/y$\updag \\
      & $\unit[160]{GeV}$ & $11.5/y$\uphashh & $1.37/y$\updag & $2.31/y$\updag \\
      \hline 
      & $\unit[30]{GeV}$ & $1.01$\upddag & $1.20$\upddag &---
      \\ 
      & $\unit[60]{GeV}$ & $1.92$\upddag & $2.72$\upddag &---
      \\ 
      & $\unit[80]{GeV}$ & $12.5$\upddag & $2.16$\updag & ---
      \\ 
\Pfivep      & $\unit[100]{GeV}$ & $18.2$\uphashh & $2.60$\updag & ---\\ 
      & $\unit[120]{GeV}$ & $19.0$\uphashh & $3.03$\updag & ---\\ 
      & $\unit[140]{GeV}$ & $26.0$\uphashh & $5.63$\updag & --- \\ 
      & $\unit[160]{GeV}$ & $71.4$\uphashh & $8.23$\updag & --- \\ \hline       
     \end{tabular} 
\end{center}
\end{table}

The minimal one-family technicolor model of Farhi and
Susskind~\cite{ef-ls} is a classic example of a technicolor model with
PNGBs.  The model contains one color singlet technilepton doublet,
\Ltechni, and one color triplet techniquark doublet, \Qtechni, while
the right-handed technifermions are all electroweak singlets.  From
Equation~\ref{eqn:ftechnisecond}, we find $f=v/2=\unit[123]{GeV}$. The
neutral PNGBs, described in terms of their technifermion quantum
numbers and normalized as in Equation~\ref{eqn:normalization} are
given by
\begin{equation}
  P^1 = \frac{1}{4\sqrt{3}}\left(3\Ltechnibar\gammafive \Ltechni -
    \Qtechnibar \gammafive \Qtechni \right)\qquad
  P^3 = \frac{1}{2\sqrt{3}}\left(3\Ltechnibar\gammafive \tau^3
    \Ltechni - \Qtechnibar \gammafive \tau^3 \Qtechni \right)\ .
\end{equation}
These PNGBs decay dominantly in the two jets mode, either to \qantiq\ 
via Extended Technicolor gauge bosons or QCD gluons, or in the case of
the \Pone, by direct decays to gluon pairs~\cite{gr-ehs, ef-ls}.
Therefore, the limits on $\NTC\anompzp$ and $\NTC\anompzz$ from
hadronic scalar decays (the $\jetj\jetj$ modes from
Tables~\ref{tab:independent} and~\ref{tab:dependent}) apply.  Because
the anomaly factors for these PNGBs (from Equations~\ref{eqn:anompzp}
and~\ref{eqn:anompzz}) are rather small,
\begin{equation}
  \label{eqn:onefamilyanoms}
  \begin{gathered}
    \anomppp^1 = \frac{1}{3\sqrt{3}} \approx 0.192 \qquad
    \anompzp^1 = \frac{1}{3\sqrt{3}} \swsq \approx 0.044 \qquad
    \anompzz^1 = \frac{1}{3\sqrt{3}} \swfour \approx 0.010 \\
    \anomppp^3 = \frac{1}{\sqrt{3}} \approx 0.577 \qquad
    \anompzp^3 = \frac{1}{4\sqrt{3}} \left(1-4\swsq\right) \approx
    0.012 \qquad
    \anompzz^3 = \frac{1}{2\sqrt{3}} \swsq \left(1-2\swsq\right)
    \approx 0.036\ .
  \end{gathered}
\end{equation}
one obtains only weak limits on the size of the technicolor group;
e.g. for $M_\Pone \leq 30$ GeV, one has $\NTC \leq 30$.  The
constraints derived from \anompzz\ are even weaker.  The results for
the light PNGB $P^0 = (3 \Etechnibar \gammafive \Etechni - \Dtechnibar
\gammafive \Dtechni)/\sqrt{24}$ in the model of Casalbuoni \textit{et.
  al.}~\cite{rc-etal} are very similar, since the anomaly factors are
equally small and $f_\pngb = \unit[123]{GeV}$.

The one-family technicolor model of Applequist and
Terning~\cite{ta-jt} includes PNGBs with $f_\pngb < v/2$.  This model
was designed as an example of a realistic technicolor scenario that
reduced the estimated technicolor contributions to the $S$ and $T$
parameters.  QCD interactions and near-critical Extended Technicolor
interactions combine to violate isospin symmetry strongly, and enhance
quark and techniquark masses relative to lepton and technilepton
masses.  In the limit of extreme isospin breaking, the techniquarks
dominate the Goldstone bosons eaten by the electroweak gauge bosons,
leaving two light, non-degenerate neutral PNGBs composed mostly of
technileptons,
\begin{equation}
  P_\Ntechni = \frac{1}{\sqrt{2}} \Ntechnibar \gammafive \Ntechni\qquad
  P_\Etechni = \frac{1}{\sqrt{2}} \Etechnibar \gammafive \Etechni\ ,
\end{equation}
with separate decay constants, $f_\Ntechni < f_\Etechni$.  The anomaly factors
for these PNGBs are not large
\begin{equation}
  \begin{gathered}
    \anomppp^\Ntechni = \anompzp^\Ntechni = 0\qquad
    \anompzz^\Ntechni = \frac{1}{8\sqrt{2}} \approx 0.088\\
    \anomppp^\Etechni = \frac{1}{\sqrt{2}} \approx 0.707\qquad
    \anompzp^\Etechni = \frac{1}{4\sqrt{2}} \left(1-4\swsq\right)
    \approx 0.014\qquad \anompzz^\Etechni = \frac{1}{8\sqrt{2}}
    \left(4 \swfour + \left(1-2\swsq\right)^2\right) \approx 0.044\ .
  \end{gathered}
\end{equation}
In the most optimistic cases where $\PEtechni \to \photon\photon$ and
$\PNtechni \to \missingE$ are the dominant decay modes, the limits
from Tables~\ref{tab:independent} and~\ref{tab:dependent} yield the
results on \NTC\ shown in Table~\ref{tab:revenge}.  Since the anomaly
factor for \PNtechni\ is so small, the limits on \PNtechni\ would be
phenomenologically relevant only if $f_\PNtechni \lesssim v/25$.  The
limits on \PEtechni\ are much stronger and, consequently, more
interesting.  For example, light \PEtechni\ with $M_\PEtechni <
\unit[60]{GeV}$ would be allowed only in models where $\NTC \leq 12$,
provided that $f_\PEtechni < v/3$.  Heavier \PEtechni, with masses in
the range from \unit[60]{GeV} to \unit[120]{GeV}, would be excluded
for $f_\PEtechni \lesssim v/6$ and would be allowed only in models
with $\NTC \lesssim 10$ even if $f_\PEtechni$ were as large as $v/2$ .

One way to obtain PNGBs with larger anomaly factors is to include
technifermions in larger representations of $\SUtwo_L$.  Manohar and
Randall created~\cite{am-lr} a one-family model with a weak isotriplet
of left-handed techniquarks, \Qtechni, of hypercharge $Y = y$ and a
weak isotriplet of left-handed technileptons, \Ltechni, of hypercharge
$Y = -3y$; the right-handed technifermions are weak singlets.  In the
absence of isospin breaking, the technipion decay constant is $f_\pngb
= v/4 = \unit[61.5]{GeV}$.  There are four neutral PNGBs, with
generators
\begin{equation}
  \begin{gathered}
    P^1 = \frac{1}{6\sqrt{2}} \left(3\Ltechnibar \gammafive \Ltechni -
      \Qtechnibar \gammafive \Qtechni\right)\qquad 
    P^3 = \frac{1}{2\sqrt{3}}\left(3\Ltechnibar \gammafive \tau^3 \Ltechni
      - \Qtechnibar \gammafive \tau^3 \Qtechni\right)\\
    P^5_{-} = \frac{1}{2\sqrt{3}} \left(3\Ltechnibar \gammafive \tau^8
      \Ltechni - \Qtechnibar \gammafive \tau^8 \Qtechni\right)\qquad
    P^5_{+} = \frac{1}{2}\left(\Ltechnibar \gammafive \tau^8 \Ltechni
      + \Qtechnibar \gammafive \tau^8 \Qtechni\right)\ ,
  \end{gathered}
\end{equation}
where $\tau^3 = \frac{1}{2} \diag(1,0,-1)$ and $\tau^8 =
\frac{1}{\sqrt{12}} \diag(1,-2,1)$.  The corresponding anomaly
factors are
\begin{equation}
  \begin{gathered}
    \anomppp^1 = 6\sqrt{2} y^2 \approx 8.485 y^2\qquad
    \anompzp^1 = 6\sqrt{2} y^2 \swsq \approx 1.948 y^2\qquad
    \anompzz^1 = 6\sqrt{2} y^2 \swfour \approx 0.449 y^2\\
    \anomppp^3 = 4 \sqrt{3} y \approx 6.928 y\qquad
    \anompzp^3 = \sqrt{3} \left(1 -4\swsq\right)y \approx
    0.139 y \qquad
    \anompzz^3 = 2 \sqrt{3} \swsq \left(1-2\swsq\right)y \approx
    0.430 y\\
    \anomppp^{5-} = \anompzp^{5-} = \anompzz^{5-} = 0\\
    \anomppp^{5+} = \frac{2}{\sqrt{3}} \approx 1.155 \qquad
    \anompzp^{5+} = \frac{1}{\sqrt{3}} \left(1-2\swsq\right) \approx
    0.312\qquad
    \anompzz^{5+} = \frac{1}{\sqrt{3}} \left(1-2\swsq
      +2\swfour\right) \approx 0.373\ .
    \label{eq:named}
  \end{gathered}
\end{equation}  
\LEP\ provides no information on \Pfivem, since this PNGB has no coupling to
the \Znaught, \photon, or \ffbar\ pairs.\footnote{The \Pfivem\ does not
  couple to a pair of neutral electroweak bosons since the anomaly factors
  vanish.  Because the \Pfivem\ and \Pfivep\ are isospin two resonances, they
  do not couple to \ffbar.  The \Pfivem\ is not stable, however, since it
  can decay via QCD gluons, technigluons or Extended Technicolor gauge
  bosons.}  For the other scalars, combining Equation~\ref{eq:named} and the
results in Tables~\ref{tab:independent} and~\ref{tab:dependent}, we find
upper bounds on the size of the technicolor group as a function of
$M_{\pngb}$ and $y$.  These limits are given in Table~\ref{tab:isotriplet}.
  
  As an example of what these results reveal about particular models,
  suppose we are interested in a theory with $\NTC = 4$ and
  techniquark hypercharge $y \sim 1$.  No matter how the \Pone\ state
  decays, the \LEP\ data imply that its mass must be greater than
  \unit[120]{GeV}.  The lower bound on the mass of the \Pthree\ state
  depends sensitively on its dominant decay mode: invisible decays
  would have been seen if \Pthree\ had a mass below \unit[120]{GeV};
  diphoton decays would have been seen if the \Pthree\ mass is below
  \unit[160]{GeV}; a \Pthree\ decaying to \bbbar\ is excuded unless
  its mass lies in the range $\unit[30]{GeV} < M_\Pthree <
  \unit[60]{GeV}$.  Finally, if the \Pfivep\ leads to two-photon final
  states, its mass must be greater than about \unit[125]{GeV}; if it
  decays to \missingE\ states, its mass must be greater than about
  \unit[70]{GeV}.  The bounds on the mass of \Pfivep\ are
  insensitive to the value of hypercharge assumed; those for the other
  PNGB loosen as the hypercharge value decreases.  The bounds become
  stricter if a larger technicolor group is chosen.

\subsection{Low-scale Technicolor Models}
\label{sec:lowscale}

\begin{table}
  \begin{center}
    \caption{Limits on the number of technicolors, 
      \NTC, and weak doublets of technifermions, $N_D$, for hadronically
      decaying PNGBs in TCSM~\cite{kl1, kl2} models as a function of
      the upper bound on the PNGB mass, from Section~\ref{sec:lowscale}.  The
      superscripted labels indicate the data used to calculate the limits:
      \dag: \anomppp\ from Table~\ref{tab:independent}; \ddag: \anompzp\ from
      Table~\ref{tab:independent}.}
\label{tab:tcsm}
\vspace{\baselineskip}
    \begin{tabular}{|c|c|c|}   \hline
      \multicolumn{1}{|c|}{$M_\technipiprime \leq$}
      &\multicolumn{2}{c|}{$\NTC\sqrt{N_D} \leq$}  \\ 
      &\multicolumn{1}{c}{$\technipiprime \to \gluon\gluon$} &
      \multicolumn{1}{c|}{$\technipiprime \to \bantib$}\\ \hline
      $\unit[30]{GeV}$ & $28$\upddag & $24$\upddag \\ 
      $\unit[60]{GeV}$ & $67$\upddag & $70$\upddag \\ 
      $\unit[80]{GeV}$ & $283$\upddag & $25$\updag \\ 
      $\unit[100]{GeV}$ & --- & $40$\updag \\ 
      $\unit[120]{GeV}$ & --- & $42$\updag \\ 
      $\unit[140]{GeV}$ & --- & $49$\updag \\ 
      $\unit[160]{GeV}$ & --- & $68$\updag \\ \hline
    \end{tabular} 
  \end{center}
\end{table}

Many modern technicolor models feature a ``walking'' technicolor
coupling to eliminate large flavor-changing neutral currents~\cite{bh,
  bh2} and separate topcolor interactions~\cite{ch, ch2} to provide
the large top quark mass.  Both innovations tend to require the
presence of a large number $N_D$ of weak doublets\footnote{While
  estimates of the $S$ and $T$ parameters in technicolor theories
  assumed to have QCD-like dynamics seem to suggest that the number of
  technifermion doublets must be small, such estimates cease to apply
  if the technicolor coupling remains strong out to the Extended
  Technicolor scale as in walking models~\cite{kl-sparam}.} of
technifermions.  For a given technicolor gauge group \SUNTC, the
number of doublets required to make the gauge coupling \gTC\ run
slowly at scales above the characteristic technicolor scale,
$\Lambda_{\mathrm{TC}}$, while remaining asymptotically free can be
estimated from the one-loop beta function:
\begin{equation}
\betaTC = - \frac{\gTC^3}{16\pi^2} \left( \frac{11}{3}
  \NTC - \frac{4}{3} N_D \right) + \cdots
\label{eq:betaone}
\end{equation}
In the models of refs.~\cite{kl-ee,kl3,kl4}, for example, $N_D \approx
10$.  Likewise, topcolor-assisted technicolor models appear to need
many doublets of technifermions to accommodate the masses of the light
fermions, the mixing between light and heavy fermions, and the
dynamical breaking of topcolor~\cite{kl1, kl-ee}.  As mentioned in
Section~\ref{sec:equations}, a large number of doublets implies a
small technipion decay constants, $f_\pngb = v/\sqrt{N_D}$.

As an example of a low-scale technicolor theory, we analyze Lane's
Technicolor Straw Man Model (TCSM)~\cite{kl1, kl2}. We assume that the
lightest technifermion doublet, composed of technileptons \TUtechni\ and
\TDtechni\ with electric charges $Q_U$ and $Q_D$ respectively, can be
considered in isolation.  Following Lane, we take $Q_U = 4/3$ and $Q_D =
1/3$, and we assume that there are two, nearly degenerate neutral mass
eigenstates, whose generators are given by
\begin{equation}
  P_{\technipi} = \frac{1}{2} \left(\TUtechnibar \gammafive \TUtechni -
    \TDtechnibar \gammafive
    \TDtechni\right)\qquad
  P_\technipiprime = \frac{1}{2} \left(\TUtechnibar \gammafive \TUtechni +
    \TDtechnibar \gammafive \TDtechni\right)\ .
\end{equation}
We can then calculate the relevant anomaly factors
\begin{equation}
  \begin{gathered}
    \anomppp^\technipi = \frac{5}{6} \approx 0.833 \qquad
    \anompzp^\technipi = \frac{5}{24}\left(1-4\swsq\right) \approx
    0.017 \qquad
    \anompzz^\technipi =
    \frac{5}{12}\swsq\left(1-2\swsq\right)\approx 0.225 \\
    \anomppp^\technipiprime = \frac{17}{18} \approx 0.944 \qquad
    \anompzp^\technipiprime = \frac{17}{18}\swsq - \frac{1}{8}\approx
    0.092\qquad
    \anompzz^\technipiprime = \frac{1}{8} - \frac{1}{4}\swsq +
    \frac{17}{18}\swfour \approx 0.117\ .
  \end{gathered}
\end{equation}
We further assume~\cite{kl2} that these PNGBs decay to jets, with
$\technipi \to \bbbar$ and $\technipiprime \to \gluon\gluon, \bbbar$
dominating.

From the limits on $\NTC\anompzz$ obtained in
Section~\ref{sec:pzz:jjmissing} for $\Znaught \to \Zstar\pngb$ with
$\pngb \to j j$ and $\Zstar \to \nunubar$, we can use the value of
$\anompzz^{\technipi}$ above, to find $\NTC \le 225/\sqrt{N_D}$ for
$M_\technipi \le \unit[30]{GeV}$.  Unfortunately, this does not
provide interesting limits even for this small $M_\technipi$.

More useful is the bound that can be obtained by combining the value
of $\anompzp^\technipiprime$ and $\anomppp^\technipiprime$ above with
the limits on $\NTC\anompzp$ and $\NTC\anomppp$ obtained for hadronic
\pngb\ decays in Tables~\ref{tab:independent} and~\ref{tab:dependent}.
For the decay modes $\technipiprime \to \gluon\gluon$ or \bantib\ we
find upper bounds on $\NTC\sqrt{N_D}$ as a function of
$M_\technipiprime$, as summarized in Table~\ref{tab:tcsm}.

To clarify the meaning of these bounds, we now consider the case where
$M_\technipiprime \leq \unit[30]{GeV}$ and the \technipiprime\ decays
primarily to \qbottom\ quarks; in this case the limit $\NTC
\sqrt{N_D} \leq 24$ applies.  As a result, for $\NTC = (4, 6, 8, 10,
12)$ the largest number of electroweak doublets of technifermions
allowed by the \LEP\ data is, respectively, $N_{D} = (36, 16, 9, 5, 4)$.
The results are very similar if the two-gluon decays of the PNGB
dominate instead.

How do these results accord with the requirements of walking
technicolor?  Based on the one-loop technicolor beta function,
\betaTC\ (Equation~\ref{eq:betaone}), a slowly running \gTC\ requires
the presence of about $11 \NTC / 4$ weak doublets of technifermions.
Then according to the \LEP\ data, walking technicolor and a very light
\technipiprime\ ($M_\technipiprime < \unit[30]{GeV}$) can coexist only
in models with $\NTC = 4 \text{ or } 6$.  A similar analysis of cases
with heavier \technipiprime\ shows that the size of the technicolor
group is restricted to $\NTC \le 6$ if $M_\technipiprime =
\unit[80]{GeV}$, loosening to $\NTC \le 12$ for a \unit[160]{GeV}
\technipiprime.  The results are similar if the 2-loop \betaTC\ 
function is used\footnote{The two loop correction to \betaTC\ 
  includes the additional term~\cite{wc,dj}
\begin{equation}
-\frac{\gTC^5}{(16\pi^2)^2} \left( \frac{34}{3}\NTC^2 - \frac{26}{3} \NTC 
N_D + 2 \frac{N_D}{\NTC}\right)\ .
\end{equation}}, even for a moderately strong technicolor coupling
$\gTC^2/4\pi \sim 1$.

\begin{table}
  \begin{center}
    \caption{Limits on the number of technicolors \NTC\ in the walking 
      technicolor model ofLane and Ramana~\cite{vl-ps, Lane:1991qh} as
      a function of the upper bound on the \PthreeL\ mass, from
      Section~\ref{sec:lowscale}.  Note that $f_\PthreeL = 41 GeV$.
      The superscripted labels indicate the data used to calculate the
      limits: \dag: \anomppp\ from Table~\ref{tab:independent}; \ddag:
      \anompzp\ from Table~\ref{tab:independent}; \hashh: \anomppp\ 
      from Table~\ref{tab:dependent}.}
\label{tab:vlps}
\vspace{\baselineskip}
    \begin{tabular}{|c|c|c|}   \hline
      \multicolumn{1}{|c|}{$M_\PthreeL \leq$}
      &\multicolumn{2}{c|}{$\NTC \leq$}  \\ 
      &\multicolumn{1}{c}{$\PthreeL \to \gamma\gamma$} &
      \multicolumn{1}{c|}{$\PthreeL \to \bantib$}\\ \hline
      $\unit[30]{GeV}$ & $3.5$\uphashh & $14.7$\upddag \\ 
      $\unit[60]{GeV}$ & $3.3$\uphashh & $42.6$\upddag \\ 
      $\unit[80]{GeV}$ & $1.4$\updag  & $3.3$\updag \\ 
      $\unit[100]{GeV}$ & $1.6$\updag & $5.1$\updag \\ 
      $\unit[120]{GeV}$ & $1.9$\updag & $5.4$\updag \\ 
      $\unit[140]{GeV}$ & $3.5$\updag & $6.2$\updag \\ 
      $\unit[160]{GeV}$ & $5.1$\updag & $8.7$\updag \\ \hline
    \end{tabular} 
  \end{center}
\end{table}

As a second example, we mention what our results imply for the walking
technicolor model of Lane and Ramana~\cite{Lane:1991qh} whose LEP II
and NLC phenomenology was studied by Lubicz and
Santorelli~\cite{vl-ps}.  To make contact with their analysis, we
follow them in taking $N_D = 9$: one color-triplet of techniquarks
($N_Q = 1$) and six color-singlets of technileptons ($N_L = 6$).  Of
the several neutral PNGBs in this model, the one whose relatively large
anomaly factors and small decay constant makes it easiest to produce
is
\begin{equation}
P^3_\Ltechni = \frac{1}{2} (\Nltechnibar \gammafive \Nltechni -
\Eltechnibar \gammafive \Eltechni)\ ,
\end{equation}
where the subscript implies a sum over all $N_L$ technilepton
doublets.  This PNGB has a decay constant $f_\PthreeL =
\unit[41]{GeV}$, and anomaly factors (in our normalization),
\begin{equation}
  \begin{gathered}
    \anomppp^\PthreeL = \frac{\sqrt{N_L}}{2} \approx 1.225 \qquad
    \anompzp^\PthreeL = \frac{\sqrt{N_L}}{8} \left(1 - 4 \swsq\right)
    \approx 0.024 \qquad 
    \anompzz^\PthreeL = \frac{\sqrt{N_L}}{4} \swsq \left(1 -
      2\swsq\right) \approx 0.076 \ , 
  \end{gathered}
\end{equation}
where the numerical factors are for $N_L = 6$. This PNGB is expected
to have a mass in the range $\unit[100 - 350]{GeV}$ \cite{vl-ps}.
Depending on the value of the ETC coupling between the PNGB and
fermions, the dominant decay of this PNGB may be into a photon pair or
\bbbar.  In Table~\ref{tab:vlps}, we show the upper bound on the
size of the technicolor group as a function of PNGB mass implied by
the results in Tables~\ref{tab:independent} and~\ref{tab:dependent}.
Apparently, if the two-photon decays dominate, the PNGB must have a
mass in excess of \unit[160]{GeV}; if the \bbbar\ decay is
preferred, the mass range \unit[80]{GeV} $\leq M_\pngb \leq$
\unit[120]{GeV} is excluded.

\section{Conclusions}
\label{sec:conclusions}

Using published analyses of data from \LEPI\ and \LEPII , we have
derived improved limits on the anomalous PNGB couplings to
$\Znaught\photon$ and the first limits on couplings to
$\Znaught\Znaught$ and $\photon\photon$.  For models in which the
PNGBs decay to photons or hadrons, the bounds on \NTC\anompzp\ are a
factor of 2-3 stronger than those previously reported~\cite{gr-ehs};
for PNGBs manifesting as missing energy, the bounds are of similar
strength but extend over a larger mass range.  As a result, it is
possible to set useful constraints on the existence of light PNGBs in
non-minimal technicolor models that have large anomalous couplings of
the PNGBs to $\Znaught\photon$, $\Znaught\Znaught$, and
$\photon\photon$ and small technipion decay constants.  For example,
the data are sensitive to light \technipiprime\ in models of low-scale
technicolor which typically include of order 10 weak doublets of
technifermions or in models with weak isotriplet technifermions.

Substantial further improvements of the limits for light \pngb\,
$M_\pngb < M_\Znaught$, will require further data collection at the
\Znaught\ pole.  Operation on the \Znaught\ resonance in the \GigaZ\ 
mode of \TESLA~\cite{GigaZ}, for example, should produce more than
$10^9$ \Znaught\ events per year of operation.  This would generate
one thousand times more data per year of running than was collected by
any one of the \LEP\ experiments.  Assuming that the limits derived by
the \LEP\ collaborations are constrained by statistics, this quantity
of data should allow improvements in the cross section limits by a
factor of 30, which would lead to an improvement of at least a factor
of five in most of our limits on both $\NTC\anompzp$ and
$\NTC\anompzz$.

The search for heavier \pngb\ can be extended in several ways.  In the short
term, analysis of the complete \LEPII\ data sample should increase the reach
of each experiment.  Combining the results from different experiments could
also give some improvement in the bounds.  In the long term, a high energy
high luminosity \epem\ collider will be able to search for PNGBs with higher
masses, larger decay constants, and smaller couplings~\cite{vl-ps}.

\vspace{24pt} 
\begin{center}\textbf{Acknowledgments}\end{center} 
\vspace{12pt} 

The authors thank R.S.~Chivukula for comments on the manuscript.
E.H.S.\ acknowledges the support of the NSF Professional Opportunities
for Women in Research and Education (POWRE) program, and the
hospitality of the Harvard University Physics Department.
\textit{This work was supported in part by the National Science
  Foundation under grant PHY-0074274, and by the Department of Energy
  under grant DE-FG02-91ER40676}.

\bibliographystyle{apsrev}
\bibliography{pngb-paper}

\begin{thebibliography}{10}
\expandafter\ifx\csname bibnamefont\endcsname\relax
  \def\bibnamefont#1{#1}\fi
\expandafter\ifx\csname bibfnamefont\endcsname\relax
  \def\bibfnamefont#1{#1}\fi
\expandafter\ifx\csname url\endcsname\relax
  \def\url#1{\texttt{#1}}\fi
\expandafter\ifx\csname urlprefix\endcsname\relax\def\urlprefix{URL }\fi
\providecommand{\bibinfo}[2]{#2}
\providecommand{\eprint}[2][]{\url{#2}}

\bibitem{am-lr}
\bibinfo{author}{\bibfnamefont{A.}~\bibnamefont{Manohar}} \bibnamefont{and}
  \bibinfo{author}{\bibfnamefont{L.}~\bibnamefont{Randall}},
  \bibinfo{journal}{Phys. Lett. B} \textbf{\bibinfo{volume}{246}},
  \bibinfo{pages}{537} (\bibinfo{year}{1990}).

\bibitem{lr-ehs}
\bibinfo{author}{\bibfnamefont{L.}~\bibnamefont{Randall}} \bibnamefont{and}
  \bibinfo{author}{\bibfnamefont{E.~H.} \bibnamefont{Simmons}},
  \bibinfo{journal}{Nucl. Phys. B} \textbf{\bibinfo{volume}{380}},
  \bibinfo{pages}{3} (\bibinfo{year}{1992}).

\bibitem{gr-ehs}
\bibinfo{author}{\bibfnamefont{G.}~\bibnamefont{Rupak}} \bibnamefont{and}
  \bibinfo{author}{\bibfnamefont{E.~H.} \bibnamefont{Simmons}},
  \bibinfo{journal}{Phys. Lett. B} \textbf{\bibinfo{volume}{362}},
  \bibinfo{pages}{155} (\bibinfo{year}{1995}), \eprint{hep-ph/9507438}.

\bibitem{vl-ps}
\bibinfo{author}{\bibfnamefont{V.}~\bibnamefont{Lubicz}} \bibnamefont{and}
  \bibinfo{author}{\bibfnamefont{P.}~\bibnamefont{Santorelli}},
  \bibinfo{journal}{Nucl. Phys. B} \textbf{\bibinfo{volume}{460}},
  \bibinfo{pages}{3} (\bibinfo{year}{1996}), \eprint{hep-ph/9505336}.

\bibitem{rc-etal}
\bibinfo{author}{\bibfnamefont{R.}~\bibnamefont{Casalbuoni}} \emph{et~al.},
  \bibinfo{journal}{Nucl. Phys. B} \textbf{\bibinfo{volume}{555}},
  \bibinfo{pages}{3} (\bibinfo{year}{1999}), \eprint{hep-ph/9809523}.

\bibitem{L3photons}
\bibinfo{author}{\bibfnamefont{M.}~\bibnamefont{Acciarri}} \emph{et~al.}
  (\bibinfo{collaboration}{\Lthree}), \bibinfo{journal}{Phys. Lett.}
  \textbf{\bibinfo{volume}{B345}}, \bibinfo{pages}{609} (\bibinfo{year}{1995}),
  \eprint{CERN-PPE/94-186}.

\bibitem{DELPHImissing}
\bibinfo{author}{\bibfnamefont{P.}~\bibnamefont{Abreu}} \emph{et~al.}
  (\bibinfo{collaboration}{\DELPHI}), \bibinfo{journal}{Z. Phys.}
  \textbf{\bibinfo{volume}{C74}}, \bibinfo{pages}{577} (\bibinfo{year}{1997}),
  \eprint{CERN-PPE/96-003}.

\bibitem{L3hadron}
\bibinfo{author}{\bibfnamefont{M.}~\bibnamefont{Acciarri}} \emph{et~al.}
  (\bibinfo{collaboration}{\Lthree}), \bibinfo{journal}{Phys. Lett. B}
  \textbf{\bibinfo{volume}{388}}, \bibinfo{pages}{409} (\bibinfo{year}{1996}),
  \eprint{CERN-PPE/96-050}.

\bibitem{OPALhadron}
\bibinfo{author}{\bibfnamefont{G.}~\bibnamefont{Alexander}} \emph{et~al.}
  (\bibinfo{collaboration}{\OPAL}), \bibinfo{journal}{Z. Phys.}
  \textbf{\bibinfo{volume}{C71}}, \bibinfo{pages}{1} (\bibinfo{year}{1996}),
  \eprint{CERN-PPE/95-193}.

\bibitem{OPALzstarp}
\bibinfo{author}{\bibfnamefont{G.}~\bibnamefont{Alexander}} \emph{et~al.}
  (\bibinfo{collaboration}{\OPAL}), \bibinfo{journal}{Phys. Lett. B}
  \textbf{\bibinfo{volume}{377}}, \bibinfo{pages}{273} (\bibinfo{year}{1996}),
  \eprint{CERN-PPE/96-019}.

\bibitem{OPAL-LEPII}
\bibinfo{author}{\bibfnamefont{G.}~\bibnamefont{Abbiendi}} \emph{et~al.},
  \bibinfo{journal}{Phys. Lett. B} \textbf{\bibinfo{volume}{465}},
  \bibinfo{pages}{303} (\bibinfo{year}{1999}), \eprint{hep-ex/9907064}.

\bibitem{DELPHI-LEPII}
\bibinfo{author}{\bibfnamefont{P.}~\bibnamefont{Abreu}} \emph{et~al.},
  \bibinfo{journal}{Phys. Lett. B} \textbf{\bibinfo{volume}{458}},
  \bibinfo{pages}{431} (\bibinfo{year}{1999}), \eprint{CERN-EP-99-058}.

\bibitem{L3-LEPII}
\bibinfo{author}{\bibfnamefont{M.}~\bibnamefont{Acciarri}} \emph{et~al.},
  \bibinfo{journal}{Phys. Lett. B} \textbf{\bibinfo{volume}{489}},
  \bibinfo{pages}{102} (\bibinfo{year}{2000}), \eprint{hep-ex/0008023}.

\bibitem{DELPHI-missing-LEPII}
\bibinfo{author}{\bibfnamefont{P.}~\bibnamefont{Abreu}} \emph{et~al.},
  \bibinfo{journal}{Eur. Phys. J.} \textbf{\bibinfo{volume}{C17}},
  \bibinfo{pages}{53} (\bibinfo{year}{2000}), \eprint{CERN-EP-2000-021}.

\bibitem{ALEPH-hz-LEPII}
\bibinfo{author}{\bibfnamefont{R.}~\bibnamefont{Barate}} \emph{et~al.},
  \bibinfo{journal}{Phys. Lett. B} \textbf{\bibinfo{volume}{466}},
  \bibinfo{pages}{50} (\bibinfo{year}{1999}), \eprint{CERN-EP-99-125}.

\bibitem{L3-hz-LEPII}
\bibinfo{author}{\bibfnamefont{M.}~\bibnamefont{Acciarri}} \emph{et~al.},
  \bibinfo{journal}{Phys. Lett. B} \textbf{\bibinfo{volume}{489}},
  \bibinfo{pages}{115} (\bibinfo{year}{2000}),
  \eprint{hep-ex/0008025,CERN-EP-2000-103}.

\bibitem{sd-sr-glk}
\bibinfo{author}{\bibfnamefont{S.}~\bibnamefont{Dimopoulos}},
  \bibinfo{author}{\bibfnamefont{S.}~\bibnamefont{Raby}}, \bibnamefont{and}
  \bibinfo{author}{\bibfnamefont{G.~L.} \bibnamefont{Kane}},
  \bibinfo{journal}{Nucl. Phys. B} \textbf{\bibinfo{volume}{182}},
  \bibinfo{pages}{77} (\bibinfo{year}{1981}).

\bibitem{je-etal}
\bibinfo{author}{\bibfnamefont{J.}~\bibnamefont{Ellis}},
  \bibinfo{author}{\bibfnamefont{M.~K.} \bibnamefont{Gaillard}},
  \bibinfo{author}{\bibfnamefont{D.~V.} \bibnamefont{Nanopoulos}},
  \bibnamefont{and} \bibinfo{author}{\bibfnamefont{P.}~\bibnamefont{Sikivie}},
  \bibinfo{journal}{Nucl. Phys. B} \textbf{\bibinfo{volume}{182}},
  \bibinfo{pages}{529} (\bibinfo{year}{1981}).

\bibitem{bh}
\bibinfo{author}{\bibfnamefont{B.}~\bibnamefont{Holdom}},
  \bibinfo{journal}{Phys. Rev. D} \textbf{\bibinfo{volume}{24}},
  \bibinfo{pages}{157} (\bibinfo{year}{1981}).

\bibitem{pdb}
\bibinfo{author}{\bibfnamefont{D.}~\bibnamefont{Groom}} \emph{et~al.}
  (\bibinfo{collaboration}{Particle Data Group}), \bibinfo{journal}{Eur. Phys.
  J.} \textbf{\bibinfo{volume}{C15}}, \bibinfo{pages}{1}
  (\bibinfo{year}{2000}), \urlprefix\url{http://pdg.lbl.gov}.

\bibitem{OPALold}
\bibinfo{author}{\bibfnamefont{R.}~\bibnamefont{Akers}} \emph{et~al.}
  (\bibinfo{collaboration}{\OPAL}), \bibinfo{journal}{Z. Phys.}
  \textbf{\bibinfo{volume}{C65}}, \bibinfo{pages}{47} (\bibinfo{year}{1995}),
  \eprint{CERN-PPE/94-105}.

\bibitem{L3old}
\bibinfo{author}{\bibfnamefont{M.}~\bibnamefont{Acciarri}} \emph{et~al.}
  (\bibinfo{collaboration}{\Lthree}), \bibinfo{journal}{Phys. Lett. B}
  \textbf{\bibinfo{volume}{346}}, \bibinfo{pages}{190} (\bibinfo{year}{1995}),
  \eprint{CERN-PPE/94-216}.

\bibitem{ta-jt}
\bibinfo{author}{\bibfnamefont{T.}~\bibnamefont{Appelquist}} \bibnamefont{and}
  \bibinfo{author}{\bibfnamefont{J.}~\bibnamefont{Terning}},
  \bibinfo{journal}{Phys. Lett. B} \textbf{\bibinfo{volume}{315}},
  \bibinfo{pages}{139} (\bibinfo{year}{1993}), \eprint{hep-ph/9305258}.

\bibitem{hhref}
\bibinfo{author}{\bibfnamefont{F.}~\bibnamefont{Berends}} \bibnamefont{and}
  \bibinfo{author}{\bibfnamefont{R.}~\bibnamefont{Kleiss}},
  \bibinfo{journal}{Nucl. Phys. B} \textbf{\bibinfo{volume}{260}},
  \bibinfo{pages}{32} (\bibinfo{year}{1985}).

\bibitem{hhguide}
\bibinfo{author}{\bibfnamefont{J.~F.} \bibnamefont{Gunion}},
  \bibinfo{author}{\bibfnamefont{H.~E.} \bibnamefont{Haber}},
  \bibinfo{author}{\bibfnamefont{G.}~\bibnamefont{Kane}}, \bibnamefont{and}
  \bibinfo{author}{\bibfnamefont{S.}~\bibnamefont{Dawson}},
  \emph{\bibinfo{title}{The Higgs Hunter's Guide}}
  (\bibinfo{publisher}{Ad\-di\-son-Wes\-ley Publishing Company},
  \bibinfo{address}{Reading, MA}, \bibinfo{year}{1990}).

\bibitem{hzref1}
\bibinfo{author}{\bibfnamefont{B.}~\bibnamefont{Ioffe}} \bibnamefont{and}
  \bibinfo{author}{\bibfnamefont{V.}~\bibnamefont{Khoze}},
  \bibinfo{journal}{Sov. J. Nucl. Phys.} \textbf{\bibinfo{volume}{9}},
  \bibinfo{pages}{50} (\bibinfo{year}{1978}).

\bibitem{hzref2}
\bibinfo{author}{\bibfnamefont{B.~E.} \bibnamefont{Lee}},
  \bibinfo{author}{\bibfnamefont{C.}~\bibnamefont{Quigg}}, \bibnamefont{and}
  \bibinfo{author}{\bibfnamefont{H.~B.} \bibnamefont{Thacker}},
  \bibinfo{journal}{Phys. Rev. D} \textbf{\bibinfo{volume}{16}},
  \bibinfo{pages}{1519} (\bibinfo{year}{1977}).

\bibitem{ef-ls}
\bibinfo{author}{\bibfnamefont{E.}~\bibnamefont{Farhi}} \bibnamefont{and}
  \bibinfo{author}{\bibfnamefont{L.}~\bibnamefont{Susskind}},
  \bibinfo{journal}{Phys. Rep.} \textbf{\bibinfo{volume}{74}},
  \bibinfo{pages}{277} (\bibinfo{year}{1981}).

\bibitem{kl1}
\bibinfo{author}{\bibfnamefont{K.}~\bibnamefont{Lane}}, \bibinfo{journal}{Phys.
  Rev. D} \textbf{\bibinfo{volume}{60}}, \bibinfo{pages}{075007}
  (\bibinfo{year}{1999}), \eprint{hep-ph/9903369}.

\bibitem{kl2}
\bibinfo{author}{\bibfnamefont{K.}~\bibnamefont{Lane}},
  \emph{\bibinfo{title}{Technihadron production and decay rates in the
  {Technicolor Straw Man Model}}} (\bibinfo{year}{1999}),
  \bibinfo{note}{companion to \cite{kl1}}, \eprint{hep-ph/9903372}.

\bibitem{bh2}
\bibinfo{author}{\bibfnamefont{B.}~\bibnamefont{Holdom}},
  \bibinfo{journal}{Phys. Lett. B} \textbf{\bibinfo{volume}{150}},
  \bibinfo{pages}{301} (\bibinfo{year}{1985}).

\bibitem{ch}
\bibinfo{author}{\bibfnamefont{C.~T.} \bibnamefont{Hill}},
  \bibinfo{journal}{Phys. Lett. B} \textbf{\bibinfo{volume}{266}},
  \bibinfo{pages}{419} (\bibinfo{year}{1991}).

\bibitem{ch2}
\bibinfo{author}{\bibfnamefont{C.~T.} \bibnamefont{Hill}},
  \bibinfo{journal}{Phys. Lett. B} \textbf{\bibinfo{volume}{345}},
  \bibinfo{pages}{483} (\bibinfo{year}{1995}), \eprint{hep-ph/9411426}.

\bibitem{kl-sparam}
\bibinfo{author}{\bibfnamefont{K.}~\bibnamefont{Lane}}, in
  \emph{\bibinfo{booktitle}{International Conference on High Energy Physics,
  27th}}, edited by \bibinfo{editor}{\bibfnamefont{P.~J.} \bibnamefont{Bussey}}
  \bibnamefont{and} \bibinfo{editor}{\bibfnamefont{I.~G.}
  \bibnamefont{Knowles}} (\bibinfo{publisher}{IOP}, \bibinfo{year}{1994}), pp.
  \bibinfo{pages}{543--548}, \eprint{hep-ph/9409304}.

\bibitem{kl-ee}
\bibinfo{author}{\bibfnamefont{K.}~\bibnamefont{Lane}} \bibnamefont{and}
  \bibinfo{author}{\bibfnamefont{E.}~\bibnamefont{Eichten}},
  \bibinfo{journal}{Phys. Lett.} \textbf{\bibinfo{volume}{B352}},
  \bibinfo{pages}{382} (\bibinfo{year}{1995}), \eprint{hep-ph/9503433}.

\bibitem{kl3}
\bibinfo{author}{\bibfnamefont{K.}~\bibnamefont{Lane}}, \bibinfo{journal}{Phys.
  Rev. D} \textbf{\bibinfo{volume}{54}}, \bibinfo{pages}{2204}
  (\bibinfo{year}{1996}), \eprint{htp-ph/9602221}.

\bibitem{kl4}
\bibinfo{author}{\bibfnamefont{K.}~\bibnamefont{Lane}}, \bibinfo{journal}{Phys.
  Lett. B} \textbf{\bibinfo{volume}{433}}, \bibinfo{pages}{96}
  (\bibinfo{year}{1998}), \eprint{hep-ph/9805254}.

\bibitem{wc}
\bibinfo{author}{\bibfnamefont{W.}~\bibnamefont{Caswell}},
  \bibinfo{journal}{Phys. Rev. Lett.} \textbf{\bibinfo{volume}{33}},
  \bibinfo{pages}{244} (\bibinfo{year}{1974}).

\bibitem{dj}
\bibinfo{author}{\bibfnamefont{D.~R.~T.} \bibnamefont{Jones}},
  \bibinfo{journal}{Nucl. Phys. B} \textbf{\bibinfo{volume}{75}},
  \bibinfo{pages}{531} (\bibinfo{year}{1974}).

\bibitem{Lane:1991qh}
\bibinfo{author}{\bibfnamefont{K.}~\bibnamefont{Lane}} \bibnamefont{and}
  \bibinfo{author}{\bibfnamefont{M.~V.} \bibnamefont{Ramana}},
  \bibinfo{journal}{Phys. Rev. D} \textbf{\bibinfo{volume}{D44}},
  \bibinfo{pages}{2678} (\bibinfo{year}{1991}).

\bibitem{GigaZ}
\bibinfo{author}{\bibfnamefont{J.}~\bibnamefont{Erler}},
  \bibinfo{author}{\bibfnamefont{S.}~\bibnamefont{Heinemeyer}},
  \bibinfo{author}{\bibfnamefont{W.}~\bibnamefont{Hollik}},
  \bibinfo{author}{\bibfnamefont{G.}~\bibnamefont{Weiglein}}, \bibnamefont{and}
  \bibinfo{author}{\bibfnamefont{P.}~\bibnamefont{Zerwas}},
  \bibinfo{journal}{Phys. Lett. B} \textbf{\bibinfo{volume}{486}},
  \bibinfo{pages}{125} (\bibinfo{year}{2000}), \eprint{hep-ph/0005024}.

\end{thebibliography}

\end{document}